\documentclass{Pos}
\usepackage{epsfig}
\usepackage{dcolumn}   
\usepackage{bm}        
\usepackage{amssymb}  

\title{Hagedorn States and Thermalization}

\ShortTitle{Hagedorn States and Thermalization}

\author{\speaker{Carsten Greiner}\\
        Goethe University Frankfurt\\
        E-mail: \email{carsten.greiner@th.physik.uni-frankfurt.de}}

\author{Jacquelyn Noronha-Hostler\\
        Frankfurt Institute for Advanced Studies\\
        E-mail: \email{hostler@th.physik.uni-frankfurt.de}}

\author{Jorge Noronha\\
        Instituto de F\'{i}sica, Universidade de S\~{a}o Paulo, S\~{a}o Paulo, SP, Brazil\\
        E-mail: \email{noronha@if.usp.br}}

\abstract{In recent years, Hagedorn states have been used to explain the equilibrium
and transport properties of a hadron gas close to the QCD critical
temperature.  These massive resonances are shown to lower $\eta/s$ to near
the AdS/CFT limit close to the phase transition. A comparison of the
Hagedorn model to recent lattice results is made and it is found that the
hadrons can reach chemical equilibrium almost immediately, well before the
chemical freeze-out temperatures found in thermal fits for a hadron gas
without Hagedorn states.}

\FullConference{ XLIX International Winter Meeting on Nuclear Physics\\
		 24-28 January 2011\\
		 BORMIO, Italy}

\begin{document}

\section{Introduction}

Rolf Hagedorn came up with the idea that the hadron mass spectrum should follow an exponential law in the 1960's \cite{Hagedorn:1968jf}.  Resonances that follow that mass spectrum are known as Hagedorn states and they become important close to the QCD critical temperature while they are exponentially suppressed at lower temperatures.  Their large masses open up the phase space for multi-particle decays.
A recent analysis of the experimental evidence for the Hagedorn spectrum can be found in \cite{Broniowski:2004yh}.  Moreover, thoughts on observing Hagedorn states in experiments are given in \cite{Bugaev:2008nu} and their usage as a thermostat in \cite{Moretto:2006zz}. Depending on the intrinsic parameters, Hagedorn states can also be used to trigger the order of the phase transition \cite{Zakout:2007nb,Zakout:2006zj}.

Bound and resonance states are due to strong interactions and all of them (including not-yet discovered ones) must be included in order to simulate all the attractive hadronic interactions \cite{Hagedorn:1994sc}. Likewise, repulsive interactions in principle must also be included and that can be done, for example, through volume corrections \cite{Hagedorn:1994sc}. According to Hagedorn, the full spectrum is then obtained by considering clusters that are formed of clusters  \cite{Hagedorn:1968jf}. Thus, Hagedorn proposed \cite{Hagedorn:1968jf} that the density of hadronic states with mass $m> 2$ GeV should be
\begin{equation}\label{eqn:fitrho}
\frac{A}{\left[m^2 +(m_{0})^2\right]^{\frac{5}{4}}}e^{\frac{m}{T_{H}}}
\end{equation}
in order to obtain the spectra from $p-p$ and $\pi-p$ scatterings. Above, $A$ is a constant, $m_0=0.5$ GeV, and $T_{H}$ is the Hagedorn temperature. At that time resonances were only known up to $\Delta (1232)$, which gave a Hagedorn temperature of $T_H\approx 160$ MeV.

Recently, we have found that Hagedorn states can account for quick dynamical chemical equilibration times within the hadron gas phase \cite{NoronhaHostler:2007jf,NoronhaHostler:2009cf,NoronhaHostler:2008zz,NoronhaHostler:2007fg,NoronhaHostler:2010dc}.  Also, Hagedorn states have been shown to contribute to the physical description of a hadron gas close to $T_c$.  The inclusion of Hagedorn states leads to a low $\eta/s$ in the hadron gas phase \cite{NoronhaHostler:2008ju,NoronhaHostler:2009hp}, which nears the supergravity bound $\eta/s=1/(4\pi)$ \cite{Kovtun:2004de}. Calculations of the trace anomaly including Hagedorn states also fits recent lattice results well and correctly describe the minimum of  the speed of sound squared, $c_s^2,$ near the phase transition found on the lattice \cite{NoronhaHostler:2008ju}.  Furthermore, it has been shown \cite{NoronhaHostler:2009tz,NoronhaHostler:2010yc} that Hagedorn states also play a (small) role in thermal model fits of hadron yield particle ratios.

\section{Strangeness Enhancement}

(Anti-)strangeness enhancement was first observed,  primarily in anti-hyperons, multi-strange baryons, and kaons, at CERN-SPS energies in comparison to p+p data.  Originally, it was considered a signature for QGP because using binary strangeness production/exchange reactions 
\begin{eqnarray}\label{eqn:strangeprod}
\pi+\bar{p}&\leftrightarrow &\bar{K}+\bar{\Lambda}\nonumber\\
K+\bar{p}&\leftrightarrow& \pi+\bar{\Lambda}
\end{eqnarray} 
chemical equilibrium could not be reached  within the hadron gas phase \cite{Koch:1986ud}.  The estimated time scale of chemical equilibration for binary reactions within a hadron gas model was $\tau\approx 1000$ fm/c \cite{Koch:1986ud}, which is significantly longer than the estimated lifetime of a hadronic fireball of about $10$ fm/c.  The abundancies after a typical hadron lifetime were somewhere between 20-100 times lower than the calculated chemical equilibrium values  \cite{Koch:1986ud}. 

Looking into the quark gluon plasma phase, the quarks and gluons can efficiently produce strange particles.  The production of $s\bar{s}$ quarks at the lowest order of perturbation QCD is through the collision of 2 gluons or the annihilation of a light anti-light quark pair.  Invariant matrix elements can be calculated, which leads to the corresponding cross-sections.  Including these cross sections in rate equations, it was found that reactions involving gluons in the deconfined phase could more quickly produce strange quarks.  Therefore, it was conjectured that strangeness enhancement was a signal for deconfinement because gluon fusion would be the primary contributor to the abundance of strange particles following hadronization and rescattering of strange quarks \cite{Koch:1986ud}.  

At the time it was assumed that multi-mesonic collisions would not play a significant role in strangeness production because their cross section would be too small.  However, Rapp and Shuryak showed \cite{Rapp:2000gy} that for SPS energies it is possible for multi-pions to interact and form anti-baryons 
\begin{equation}\label{eqn:antiproton}
\bar{p}+N\leftrightarrow n\pi,
\end{equation}
which has a cross-section of $\sigma_{\bar{p}N}\approx 50\;\mathrm{mb}$.  Using rate equations, one finds that the chemical equilibration time is proportional to the inverse of the thermal reaction rate
\begin{equation}\label{eqn:spstau}
\tau_{\bar{p}}=\frac{1}{\langle\langle\sigma_{\bar{p}+N\leftrightarrow n\pi}v_{\bar{p}N}\rangle\rangle\rho_{N}}\approx 1-3 \;fm/c
\end{equation}
where $v_{\bar{p}N}\approx0.5c$ and the baryonic density is $\rho_{N}\approx \rho_{0}\;\mathrm{to}\;2\rho_{0}$, which is typical for SPS.
Greiner and Leupold \cite{Greiner} extended this idea to anti-hyperons
\begin{equation}\label{eqn:antihyp}
\bar{Y}+N\leftrightarrow  n\pi+n_{Y}K,
\end{equation}
which also gives time scales on the order of Eq.\ (\ref{eqn:spstau}). 
Therefore, due to multi-mesonic collisions, the chemical 	equilibration time scales are
short enough to account for chemical equilibration within a cooling hadronic
fireball at SPS energies.

A problem arises if we use the same multi-mesonic reactions in the hadron gas phase at RHIC temperatures where experiments show that the particle abundances reach chemical equilibration close to the phase transition \cite{Braun-Munzinger}.  At RHIC, assuming $T=170$ MeV, we use Eq.\ (\ref{eqn:spstau}) where $\sigma\approx
30\;\mathrm{mb}$ and 
$\rho_{B}^{eq}=\rho_{\bar{B}}^{eq}\approx0.04\;\mathrm{fm}^{-3}$ (Note that at RHIC there is approximately an equal number of baryons and anti-baryons \cite{WhitePapers}. Additionally, the density can be calculated within a grand canonical model.), and find that  the equilibrium rate of $\Omega$  is
$\tau_{\Omega}\approx
10\;\frac{\mathrm{fm}}{\mathrm{c}}$, which is considerably longer than the fireball's lifetime of $\tau<
4\;\frac{\mathrm{fm}}{\mathrm{c}}$ in the hadronic stage.  Moreover, $\tau_{\Omega}\approx
10\;\frac{\mathrm{fm}}{\mathrm{c}}$ was also obtained in \cite{Kapusta} using the fluctuation-dissipation theorem
and Ref.\ \cite{Huovinen:2003sa} found thrice lower populations than experiments for various anti-hyperons in the $5\%$ most central Au+Au collisions (also see \cite{BSW}). 
These discrepancies led to the suggestion that the hadrons are ``born" into equilibrium i.e.
the system is already in a chemically frozen out state at the end of
the phase transition \cite{Stock:1999hm,Heinz:2006ur}.

\subsection{Model}\label{model}

Hagedorn states are massive resonances that have large decay widths, which can open up the phase space to multi-particle collisions.  Because Hagedorn states decay so quickly they can catalyse quick reactions between hadrons that would otherwise have smaller cross-sections and take longer to reach chemical equilibrium.  These reactions follow the general form  
\begin{equation}\label{eqn:decay}
n\pi\leftrightarrow HS\leftrightarrow n\pi+X\bar{X}
\end{equation}
where $X\bar{X}=p\bar{p}$, $K\bar{K}$, $\Lambda\bar{\Lambda}$, or $\Omega\bar{\Omega}$.  Our idea is that these very massive Hagedorn states exist and are so large that they decay almost immediately into multiple pions and $X\bar{X}$ pairs.   We note that in this work we consider only non-strange, mesonic Hagedorn states.

The exponential in Eq.\ (\ref{eqn:fitrho}) arises from Hagedorn's original idea that there is an exponentially growing mass spectrum. Thus, as $T_H$ is approached, Hagedorn states become increasingly more relevant and heavier resonances ``appear".  The factor in front of the exponential may appear in various forms \cite{Broniowski:2004yh,Moretto:2006zz}. While the choice in this factor can vary, it was found in \cite{Broniowski:2004yh} that the present form gives lower values of $T_H$, which more closely match the predicted lattice critical temperature \cite{Cheng:2007jq,Aoki:2005vt,Bazavov:2009zn}.   Further discussion on the parameters can be found in \cite{chatterjee,Majumder:2010ik}.

Returning to Eq.\ (\ref{eqn:fitrho}), we assume that $T_H=T_c$, and then we consider the two different different lattice results for $T_c$: $T_c=196$ MeV \cite{Cheng:2007jq,Bazavov:2009zn} (Hot Quarks collaboration), which uses an almost physical pion mass, and $T_c=176$ MeV \cite{Aoki:2005vt} (BMW collaboration). Note that there have been updated lattice results for the lower temperature region \cite{Borsanyi:2010cj} that we have yet to publish results on but has been discussed in \cite{Majumder:2010ik}. Furthermore, we need to take into account the repulsive interactions and, therefore, we use the volume corrections in \cite{NoronhaHostler:2008ju,Kapusta:1982qd,Rischke:1991ke}. 

\begin{figure}
\begin{minipage}{0.5\linewidth}
\vspace{0pt}
\centering
\epsfig{file=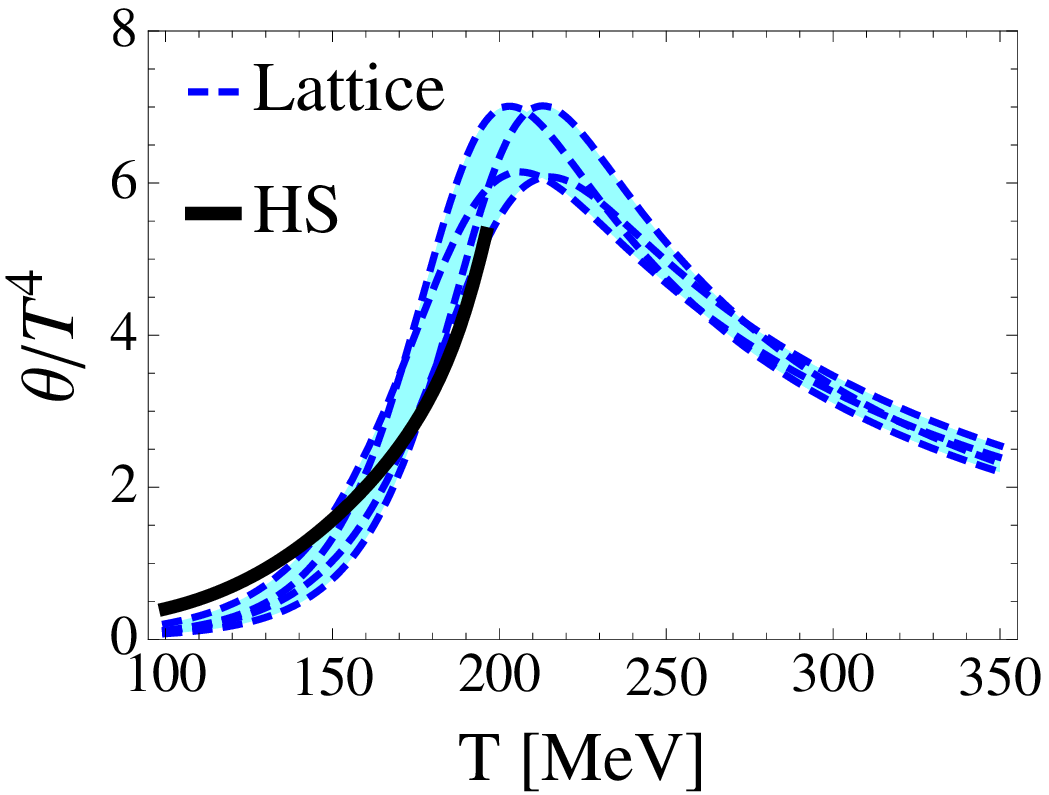,width=1\linewidth,clip=}\\
\caption{Comparison between the trace anomaly using a hadron resonance gas model with Hagedorn states \cite{NoronhaHostler:2007jf} (solid black line) and without (black dashed line) \cite{StatModel}. Lattice data points for the $p4$ action with $N_{\tau}=6$ \cite{Cheng:2007jq} are also shown. }
\label{fig:eT4}
\end{minipage}
\hspace{0.5cm}
\begin{minipage}{0.5\linewidth}
\vspace{0pt}
\centering
\epsfig{file=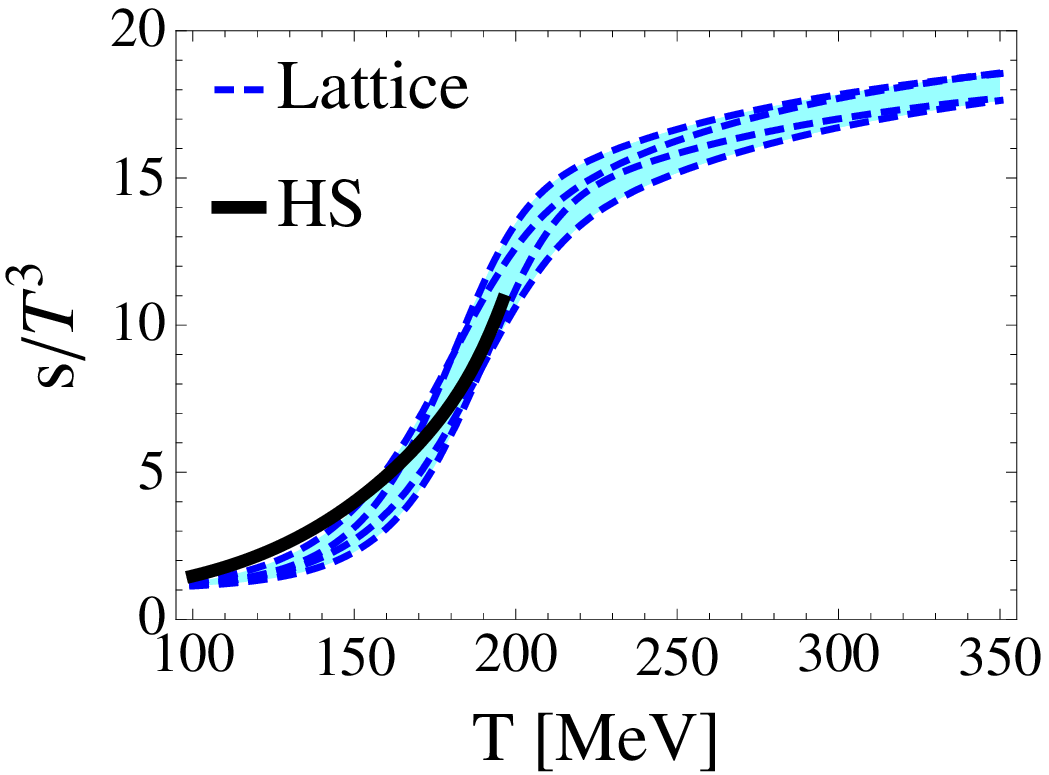,width=1\linewidth,clip=}\\~\\
\caption{ Comparison of entropy density to lattice QCD results from  \cite{Cheng:2007jq,Bazavov:2009zn} where $T_c=196$ MeV. HS is in reference to our model including Hagedorn states.   }
\label{fig:sT4}
\end{minipage}
\end{figure}

In order to find the maximum Hagedorn state mass $M$ and the ``degeneracy" A, we fit our model to the thermodynamic properties of the lattice.  
In the RBC-Bielefeld collaboration the thermodynamical properties are derived from 
the quantity $\varepsilon-3p$, the so-called
interaction measure, which is what we fit in order to obtain the parameters for the Hagedorn states.  Thus, we obtain $T_H=196$ MeV, $A=0.5 {\rm GeV}^{3/2}$, $M=12$ GeV, and $B=\left(340 {\rm MeV}\right)^4$.  The fit for the trace anomaly $\Theta/T^4$ is shown in Fig.\ \ref{fig:eT4}.  We also show the fit for the entropy density in Fig.\ \ref{fig:sT4}.  Both fits are within the lattice error bars and mimic the behavior of the lattice results.  As discussed in \cite{NoronhaHostler:2008ju}, a hadron resonance gas model with Hagedorn states is able to fit the lattice data whereas a hadron resonance gas without Hagedorn states (but with excluded volume corrections) misses the general behavior displayed by the lattice data at high temperatures. Here we follow Hagedorn's idea \cite{Hagedorn:1994sc} and do not neglect the repulsive interactions between the hadrons.  

\begin{figure}
\begin{minipage}{0.5\linewidth}
\vspace{0pt}
\centering
\epsfig{file=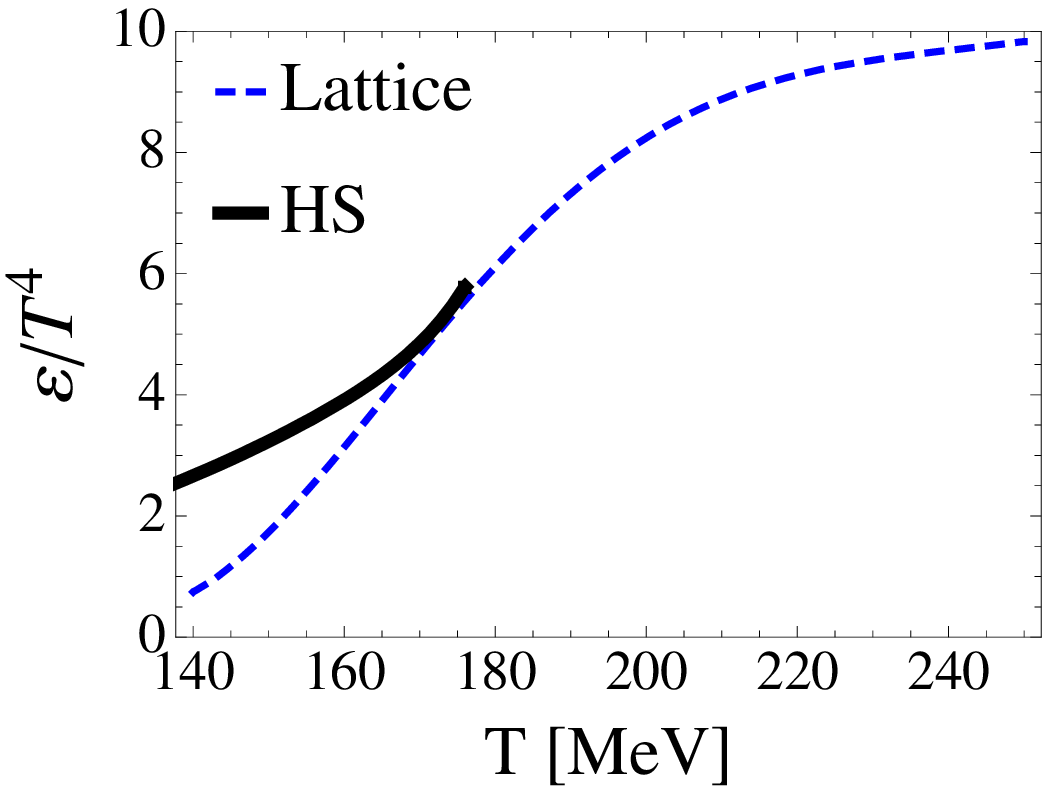,width=1\linewidth,clip=}
\caption{Comparison of energy density to lattice QCD results from  \cite{Aoki:2005vt} where $T_c=176$ MeV.  HS is in reference to our model including Hagedorn states.  }
\label{fig:eT4_173}
\end{minipage}
\hspace{0.5cm}
\begin{minipage}{0.5\linewidth}
\vspace{0pt}
\centering
\epsfig{file=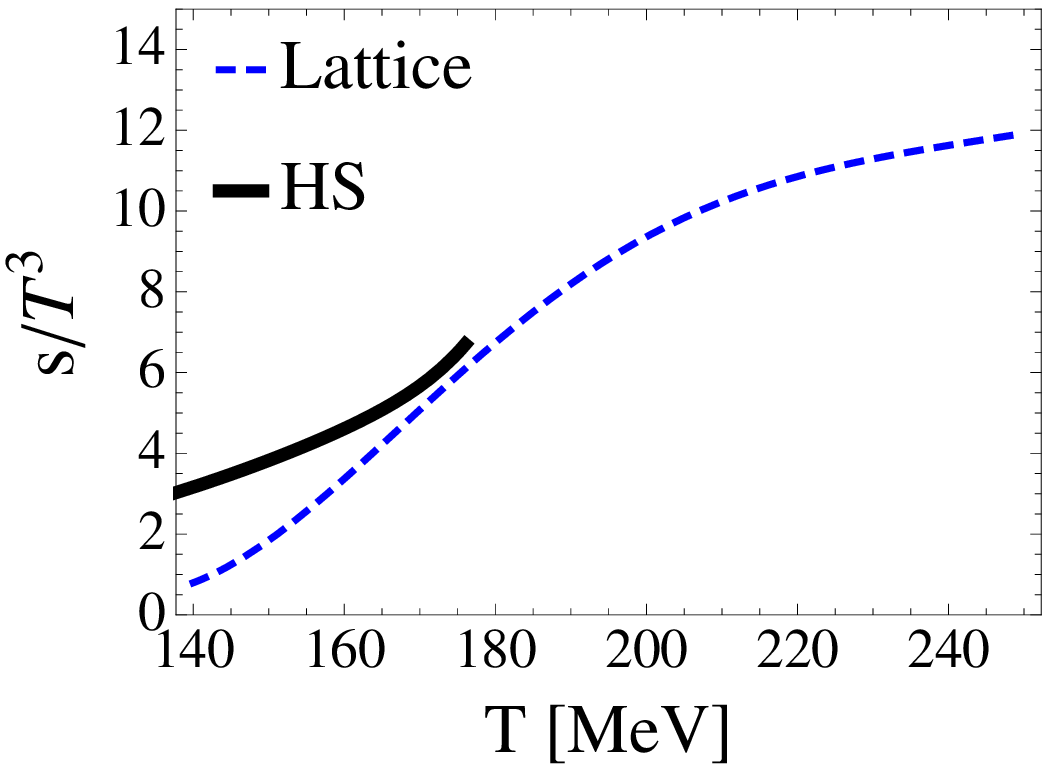,width=1\linewidth,clip=}
\caption{Comparison of entropy density to lattice QCD results from \cite{Aoki:2005vt} where $T_c=176$ MeV. HS is in reference to our model including Hagedorn states.  }
\label{fig:sT4_173}
\end{minipage}
\end{figure}

BMW obtains the thermodynamical properties differently than RBC-Bielefeld and, therefore, we fit only the energy density as shown in Fig.\ \ref{fig:eT4_173}.  From that we obtain $T_H=176$ MeV, $A=0.1 {\rm GeV}^{3/2}$, $M=12$ GeV, and $B=\left(300 {\rm MeV}\right)^4$. We also show a comparison to the entropy density in Fig.\ \ref{fig:sT4_173}  
Our results with the inclusion of Hagedorn states are able to match lattice data near the critical temperature but do not match as well at lower temperatures in Fig.\ \ref{fig:eT4} and  Fig.\ \ref{fig:sT4}. For a detailed discussion of hadron gas models and their ability to match lattice data see \cite{Huovinen:2009yb}.   

\subsubsection{Master Equations}

In order to describe the dynamics of Eq.\ (\ref{eqn:decay}) we use master equations.  They include both the forward and back reactions, which ensures that detailed balance is mantained, and the state of chemical equilibrium is a fixed point of the rate equations. Additionally, master equations are naturally suited to describe multi-particle reactions whereas the transport equations used in UrQMD \cite{urqmd} are better suited to describe the dynamics of binary collisions.  Furthermore, the Hagedorn spectrum is discretized by using mass bins of 100 MeV.  Each bin is described by its own rate equation.

The rate equations for the Hagedorn resonances $N_{i}$, pions $N_{\pi}$, and the $X\bar{X}$ pair $N_{X\bar{X}}$
\begin{eqnarray}\label{eqn:setpiHSXX}
\dot{N}_{i}&=&\Gamma_{i,\pi}\left[N_{i}^{eq}\sum_{n} B_{i,n}
\left(\frac{N_{\pi}}{N_{\pi}^{eq}}\right)^{n}-N_{i}\right]+\Gamma_{i,X\bar{X}}\left[ N_{i}^{eq}
\left(\frac{N_{\pi}}{N_{\pi}^{eq}}\right)^{\langle n_{i,x}\rangle} \left(\frac{N_{X\bar{X}}}{N_{X\bar{X}}^{eq}}\right)^2 -N_{i}\right]\nonumber\\
\dot{N}_{\pi }&=&\sum_{i} \Gamma_{i,\pi}  \left[N_{i}\langle n_{i}\rangle-N_{i}^{eq}\sum_{n}
B_{i, n}n\left(\frac{N_{\pi}}{N_{\pi}^{eq}}\right)^{n} \right]+\sum_{i} \Gamma_{i,X\bar{X}} \langle n_{i,x}\rangle\left[N_{i}-
N_{i}^{eq}
\left(\frac{N_{\pi}}{N_{\pi}^{eq}}\right)^{\langle n_{i,x}\rangle} \left(\frac{N_{X\bar{X}}}{N_{X\bar{X}}^{eq}}\right)^2\right]  \nonumber\\
\dot{N}_{X\bar{X}}&=&\sum_{i}\Gamma_{i,X\bar{X}}\left[ N_{i}- N_{i}^{eq}\left(\frac{N_{\pi}}{N_{\pi}^{eq}}\right)^{\langle n_{i,x}\rangle} \left(\frac{N_{X\bar{X}}}{N_{X\bar{X}}^{eq}}\right)^2\right].
\end{eqnarray}
The decay widths for the $i^{th}$ resonance are $\Gamma_{i,\pi}$ (for the decay into multiple pions) and $\Gamma_{i,X\bar{X}}$ (for the decay into multiple pions and a $X\bar{X}$ pair), the branching ratio is $B_{i,n}$ (clearly $\sum_{n}B_{i,n}=1$), and the average number of pions that each resonance will decay into is $\langle n_{i}\rangle$.  The equilibrium values $N^{eq}$ are both temperature and chemical potential dependent.  However, here we set $\mu_b=0$, which is a good approximation for collisions at large $\sqrt{s}$.


\subsubsection{Branching Ratios}

The branching ratio, $B_{i,n}$, is the probability that the $i^{th}$ Hagedorn state will decay into $n$ pions.  Since we are dealing with probabilities, then $\sum_{n}B_{i,n}=1$ must always hold.  We have the condition that each Hagedorn resonance must decay into at least 2 pions.  Thus, our average number of pions must be normalized to ensure that $n\geq 2$ and $\sum_{n}B_{i,n}=1$.  
The branching ratios for the reaction $HS\leftrightarrow n\pi$ are assuemd to follow the Gaussian distribution
\begin{equation}
B_{i, n}\approx
\frac{1}{\sigma_{i}\sqrt{2\pi}}e^{-\frac{(n-\langle n_{i}\rangle)^{2}}{2\sigma_{i} ^{2}}}\,,
\end{equation}
which has its peak centered at $\langle n_{i}\rangle$ and the width of the distribution is $\sigma^2$.

Assuming a statistical, microcanonical branching for the decay, we can take a linear fit to the average number of pions in
Ref.\ \cite{Greiner:2004vm} to find $\langle n_{\pi}\rangle$ such that $\langle n_i\rangle=0.9+1.2\frac{m_{i}}{m_{p}}$ is the average pion number that each
Hagedorn state decays into where $m_p$ is the mass of the proton.  In the microcanonical model the volume is $V=M_{i}/\varepsilon$ where $\varepsilon$ is the mean energy density of a Hagedorn state (taken as $\varepsilon=0.5\frac{GeV}{fm^3}$). Further discussions regarding this can be found in \cite{Greiner:2004vm,Liu}.
The width of the distrubtion is $\sigma^{2}_{i}=(0.5\frac{m_{i}}{m_{p}})^{2}$, which roughly matches the canonical description in \cite{Becattini:2004rq}. 
We have the condition that each Hagedorn resonance must decay into at least 2 pions.  Thus, after we normalize for the cutoff $n\geq 2$, we have $\langle n_{i}\rangle\approx 3 - 34$ and $\sigma_{i}^2\approx0.8 - 510$. For the average number of pions when a $X\bar{X}$ pair is present, we again refer to the micro-canonical model in \cite{Greiner:2004vm,Liu}
\begin{equation}\label{eqn:nfit}
    \langle n_{i,x}\rangle=\left(\frac{2.7}{1.9}\right)\left(0.3+0.4m_i\right)\approx 2-7.
\end{equation}
where $m_i$ is in GeV. In this paper we do not consider a distribution but rather only the average number of pions when a $X\bar{X}$ pair is present.  We  assume that $\langle n_{i,x}\rangle=\langle n_{i,p}\rangle=\langle n_{i,k}\rangle=\langle n_{i,\Lambda}\rangle=\langle n_{i,\Omega}\rangle$ for when a proton anti-proton pair, kaon anti-kaon pair, $\Lambda\bar{\Lambda}$, or  $\Omega\bar{\Omega}$ pair is present.


\subsubsection{Decay Width}

We used a linear fit for the decay width considering only the light, non-strange, mesonic resonances up to $M=2$ GeV given in \cite{Eidelman:2004wy} and exclude $f_0(600)$ because it is an extreme outlier
\begin{equation}
\label{HSdecaywidth}
\Gamma_{i} [GeV] =0.15m_{i} [GeV]-0.03,
\end{equation}
which ranges from $\Gamma_{i}=250\;\mathrm{MeV\;to}\;1800$ MeV.  The total decay width has been separated into two parts in Eq.\ (\ref{eqn:setpiHSXX}): one for the reactions $HS\leftrightarrow n\pi$, $\Gamma_{i,\pi}$, and one for the reaction $HS\leftrightarrow n\pi+X\bar{X}$, $\Gamma_{i,X\bar{X}}$, whereby $\Gamma_{i}=\Gamma_{i,\pi}+\Gamma_{i,X\bar{X}},$
which ensures that Eqs.\ (\ref{eqn:setpiHSXX}) are zero
at equilibrium. Then, the relative decay width $\Gamma_{i,X\bar{X}}$ modeled after
the decay width in reference Ref.\ \cite{Greiner:2004vm} is the average number of $X\bar{X}$ in the system $\langle X\rangle$ multiplied by the total decay width $\Gamma_{i}$,
\begin{equation}
\Gamma_{i,X\bar{X}}=\langle X\rangle\;\Gamma_{i}.
\end{equation}
That means that $\Gamma_{i,\pi}$ is then
\begin{eqnarray}
\Gamma_{i,\pi}&=&(1-\langle X\rangle)\Gamma_{i}.
\end{eqnarray}
The $\langle X\rangle$ taken from a micro-canonical model for the protons and kaons \cite{Liu,Greiner:2004vm}  and our own canonical model for the lambdas and omegas \cite{Max} such that 
\begin{eqnarray}\label{eqn:gamfit}
  p &=& 0.058\;m_{i}-0.10 \approx 0.01 - 0.6\nonumber\\
  K^{+}&=&0.075\;m_{i}+0.047\approx 0.2 - 0.95
\end{eqnarray}
and the  decay widths of $\Lambda$ and $\Omega$ are $\Gamma_{i,\Lambda\bar{\Lambda}}=3-250$ MeV and  $\Gamma_{i,\Omega\bar{\Omega}}=0.01-4$ MeV (see \cite{Max}).


\subsubsection{Initial Conditions}
The equilibrium values are found using a statistical model \cite{StatModel} with  the light and strange particles from the PDG \cite{Eidelman:2004wy} and also including effective particles from resonances.  Throughout this work our initial conditions  at $t_0$ (the point of the phase transition into the hadron gas phase) are

\begin{equation}
\label{initcond}
\alpha\equiv \frac{N_{\pi}}{N_{\pi}^{eq}}(t_0) \, , \, \beta_{i}\equiv\frac{N_{i}}{N_{i}^{eq}}(t_0) \, , \mbox{and}\,   \phi\equiv\frac{N_{X\bar{X}}}{N_{X\bar{X}}^{eq}}(t_0) \, \, , 
\end{equation}
which are chosen by holding the contribution to the total entropy from the Hagedorn states and pions constant, i.e.,  
\begin{eqnarray}\label{entrcont}
s_{Had}(T_{0},\alpha)V(t_{0})+s_{HS}(T_{0},\beta_{i})V(t_{0})=s_{Had+HS}(T_{0})V(t_{0})=const.
\end{eqnarray} 
and the corresponding initial condition configurations are shown in Tab.\ \ref{tab:IC}. $s_{Had}(T_{0},\alpha)$ is the entropy density at the initial temperature, i.e., the critical temperature multiplied by our choice in $\alpha$.  Because the hadron gas is dominated by pions we can assume that $\alpha$ represents the initial fraction of pions in equilibrium.  $s_{HS}(T_{0},\beta_{i})$ represents the entropy contribution from the Hagedorn states at $T_c$ multiplied by the initial fraction of Hagedorn states in equilibrium.  We hold $\alpha$ as a constant and then find the appropriate $\beta_i$.

\begin{table}
\begin{center}
 \begin{tabular}{|c|c|c|c|}
 \hline
 & & & \\
   & $\alpha=\lambda_{\pi}(t_0)$ & $\beta_{i}=\lambda_i(t_0)$ & $\phi=\lambda_{X\bar{X}}(t_0)$ \\
    & & & \\
 \hline
$IC_1$ & 1 & 1 & 0 \\
$IC_2$ & 1 & 1 & 0.5 \\
$IC_3$ & 1.1 & 0.5 & 0 \\
$IC_4$ & 0.95 & 1.2 & 0 \\
 \hline
 \end{tabular}
 \end{center}
 \caption{Initial condition configurations, recalling  Eq.\ (\protect\ref{initcond}).}\label{tab:IC}
 \end{table}

\subsubsection{Expansion}\label{sec:bexpan}

In order to include the cooling of the fireball we need to find a relationship between the temperature and the time, i.e., $T(t)$.  Thus, we apply a Bjorken expansion where the total entropy is held constant
\begin{equation}\label{eqn:constrain}
\mathrm{const.}=s(T)V(t)\sim\frac{S_{\pi}}{N_{\pi}}\int \frac{dN_{\pi}}{dy} dy.
\end{equation}
where $s(T)$ is the entropy density of the hadron gas with volume corrections. 
The total number of pions in the $5\%$ most central collisions $\sum_{i}N_{\pi^{i}}=\int_{-0.5}^{0.5} \frac{dN_{\pi}}{dy} dy=874$ can be found from experimental
results in \cite{Bearden:2004yx}.   The $S_{\pi}/N_{\pi}$ in our model is $\approx 6$ (see \cite{Greiner:1993jn}).

The effective volume at mid-rapidity can be parametrized as a function of time.  We do this by using a Bjorken expansion and including accelerating radial flow
\begin{equation}\label{eqn:bjorken}
V(t)=\pi\;ct\left(r_{0}+v_{0}(t-t_{0})+\frac{1}{2}a_{0}(t-t_{0})^2 \right)^2
\end{equation}
where the initial radius is $r_{0}(t_0)=7.1$ fm. For $T_H=196$ we have $t_{0}^{(196)}\approx2 fm/c$ and for $T_H=176$ we allow for a longer expansion before the hadron gas phase is reached and, thus, calculate the appropriate $t_0^{(176)}$ from the expansion starting at $T_H=196$, which is $t_0^{(176)}\approx 4 fm/c$. The $T(t)$ relation has almost no effect on the results (see \cite{NoronhaHostler:2009cf}). Therefore, we choose $v_0/c=0.5 $ and $a_0/c^2=0.025 \,{\rm fm}^{-1}$  for the remainder of this work. 

\subsubsection{Effective Numbers}

\begin{figure}
\begin{minipage}{0.5\linewidth}
\vspace{0pt}
\centering
\epsfig{file=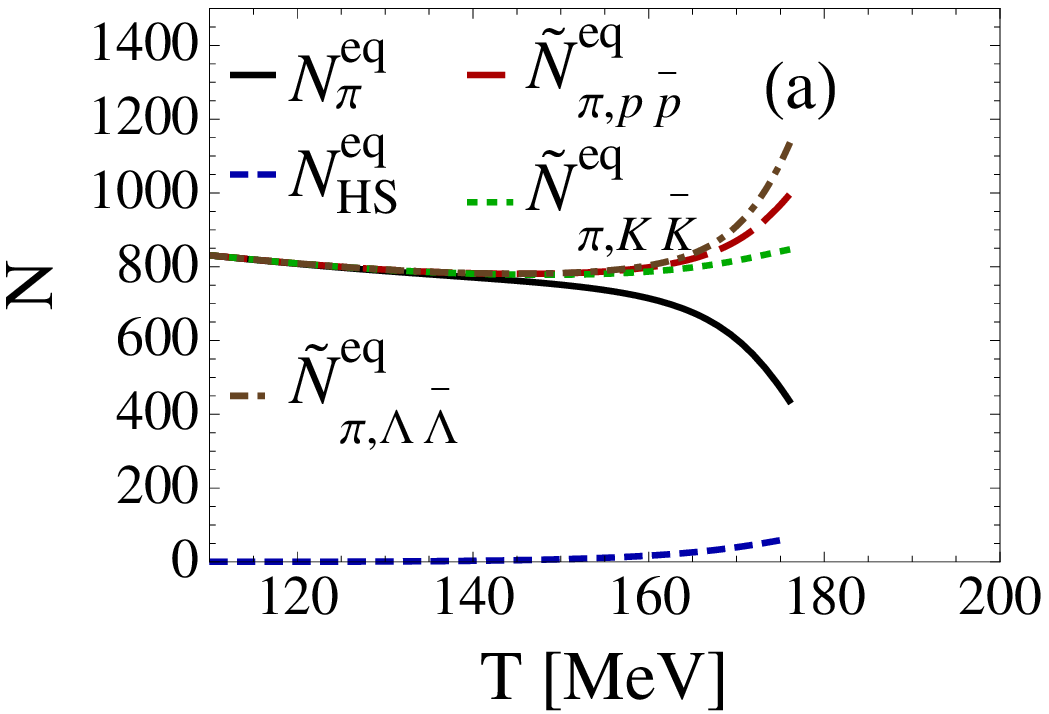,width=1\linewidth,clip=}
\end{minipage}
\hspace{0.5cm}
\begin{minipage}{0.5\linewidth}
\vspace{0pt}
\centering
\epsfig{file=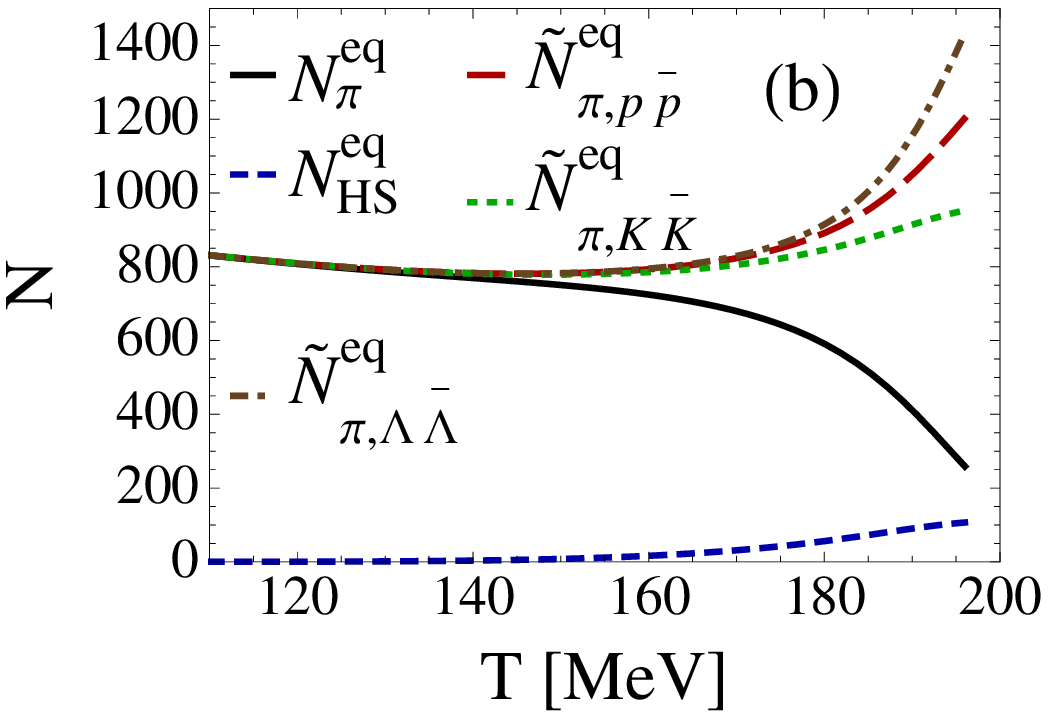,width=1\linewidth,clip=}
\end{minipage}
\caption{Comparison of the effective pion numbers when (a) $T_H=176$ MeV or (b) $T_H=196$ MeV.} \label{fig:effnum}
\end{figure}

\begin{figure}
\begin{minipage}{0.5\linewidth}
\vspace{0pt}
\centering
\epsfig{file=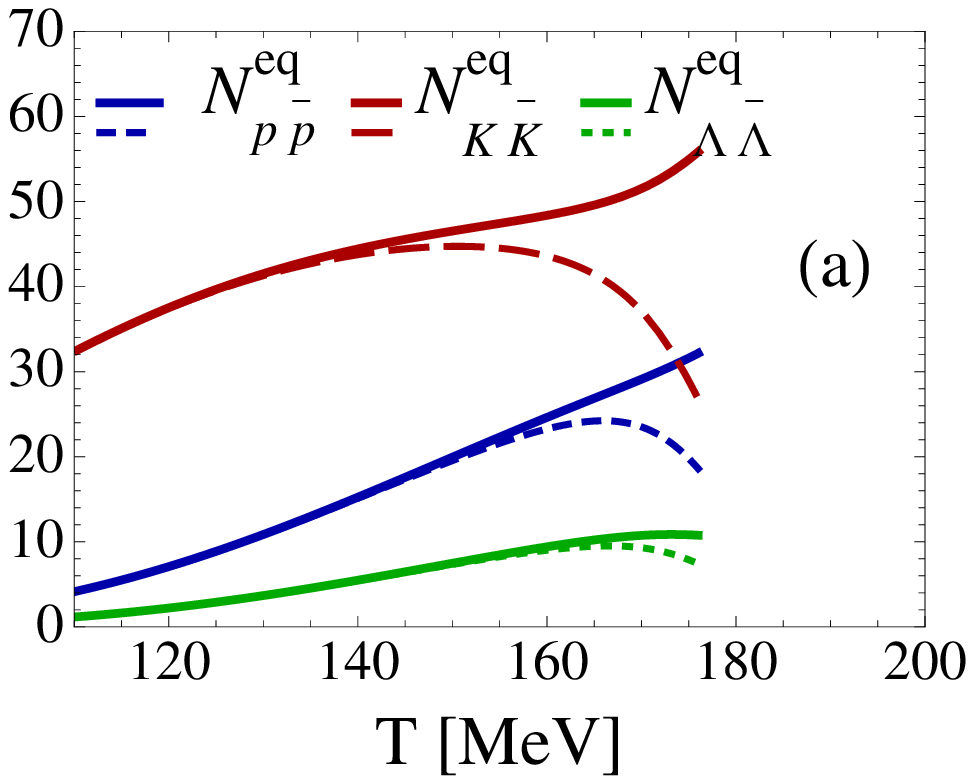,width=1\linewidth,clip=}
\end{minipage}
\hspace{0.5cm}
\begin{minipage}{0.5\linewidth}
\vspace{0pt}
\centering
\epsfig{file=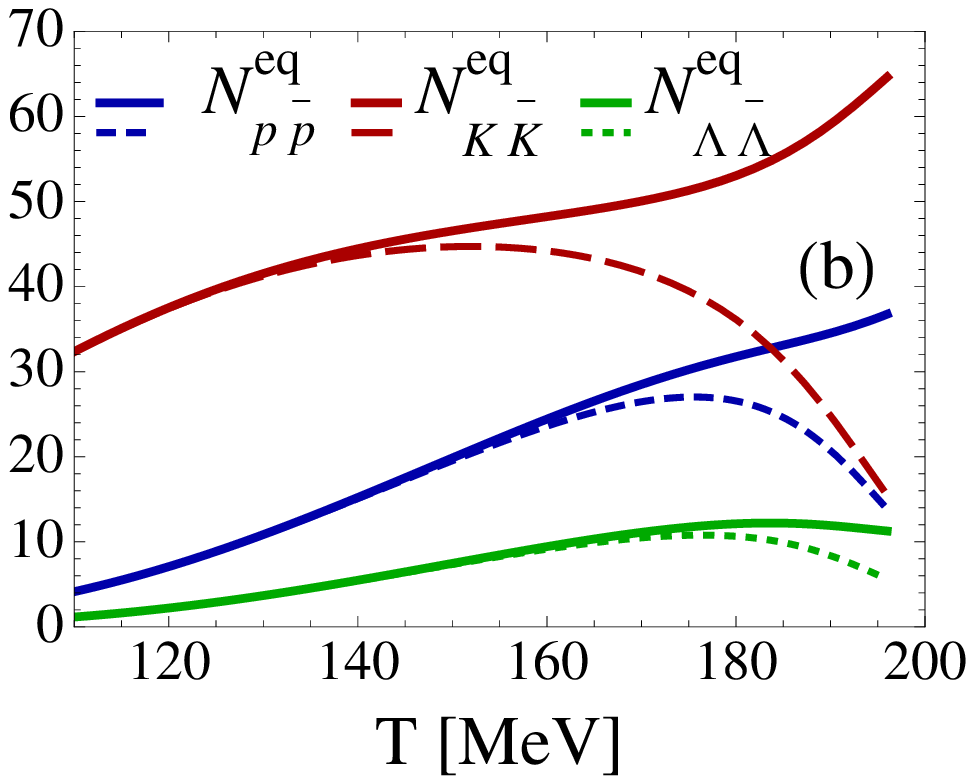,width=1\linewidth,clip=}
\end{minipage}
\caption{Comparison of the total number of $X\bar{X}$ and their effective numbers when (a) $T_H=176$ MeV  or (b) $T_H=196$ MeV.} \label{fig:density}
\end{figure}

Because the volume expansion depends on the entropy according to Eq.\ (\ref{eqn:constrain}) and the Hagedorn resonances contribute strongly to the entropy only close to the critical temperature, the equilibrium values actually decrease with increasing temperature close to $T_{c}$ for the hadrons as seen in Fig.\ \ref{fig:effnum} and Fig.\ \ref{fig:density}.  Therefore, one has to include the potential contribution of the Hagedorn resonances to the pions as in the case of standard hadronic resonances, e.g. a $ \rho $-meson decays dominantly into two pions and, thus, accounts for them by a factor two.  Including the Hagedorn state contribution, we arrive at our effective number of pions
\begin{eqnarray}\label{eqn:effpi}
\tilde{N}_{\pi,X\bar{X}}&=&N_{\pi}+\sum_{i}N_{i}\left[\left(1-\langle X_i\rangle\right)\langle n_{i}\rangle +\langle X_i\rangle\langle n_{i,x}\rangle\right]
\end{eqnarray}
which are shown in Fig.\ \ref{fig:effnum}. In  Fig.\ \ref{fig:effnum} we see that after the inclusion of the effective pion numbers that the number of pions only decreases with decreasing temperature.  Furthermore, in Fig.\ \ref{fig:effnum} the total number of Hagedorn states, $\sum_{i}N_i^{eq}$ is also shown.  While there are 
by far
fewer Hagedorn states present than pions, we see that they are important because of their large contribution to the entropy density.


Moreover, it is useful to consider the effective number of $X\bar{X}$ pairs
\begin{eqnarray}\label{eqn:effbbkk}
\tilde{N}_{X\bar{X}}&=&N_{X\bar{X}}+\sum_{i}N_{i}\langle X_i\rangle
\end{eqnarray}
because Hagedorn states also contribute strongly to the $X\bar{X}$ pairs close to $T_c$ as seen in Fig. \ref{fig:density}. Again we see that only the effective number of $X\bar{X}$ pairs have consistent decreasing behaviour with decreasing temperature whereas without the Hagedorn state contributions we see a decrease close to $T_c$.

\subsection{Results}

\begin{figure}
\begin{minipage}{0.5\linewidth}
\vspace{0pt}
\centering
\epsfig{file=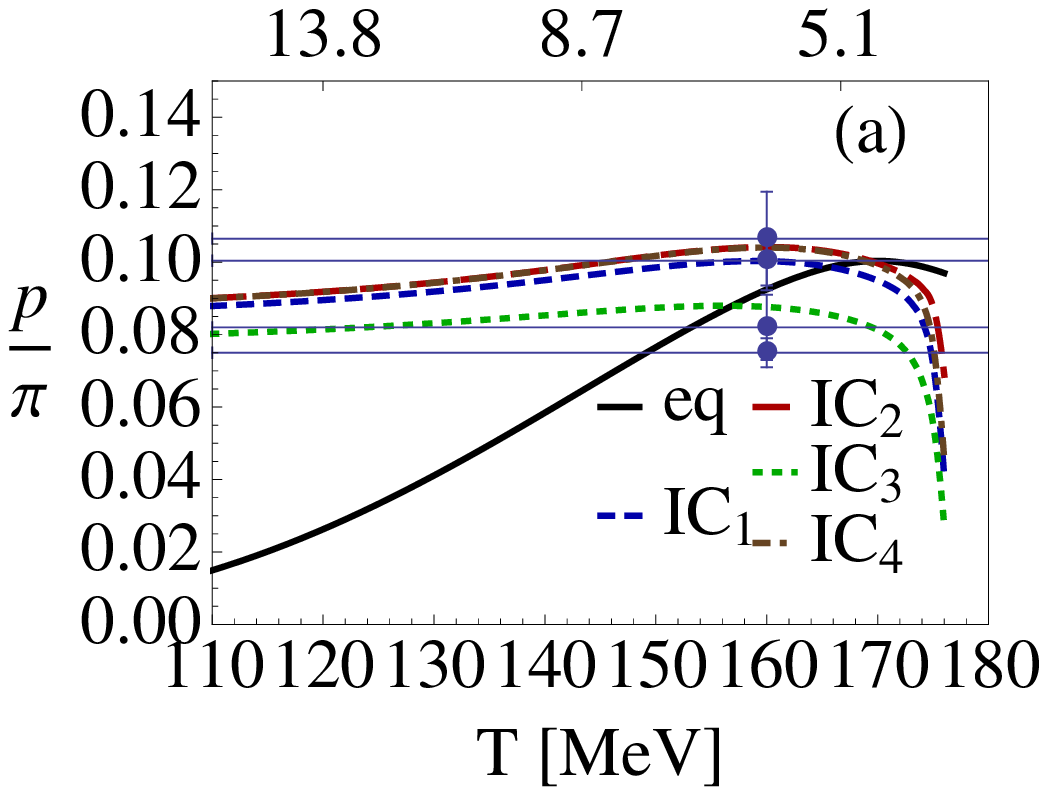,width=1\linewidth,clip=} 
\end{minipage}
\hspace{0.5cm}
\begin{minipage}{0.5\linewidth}
\vspace{0pt}
\centering
\epsfig{file=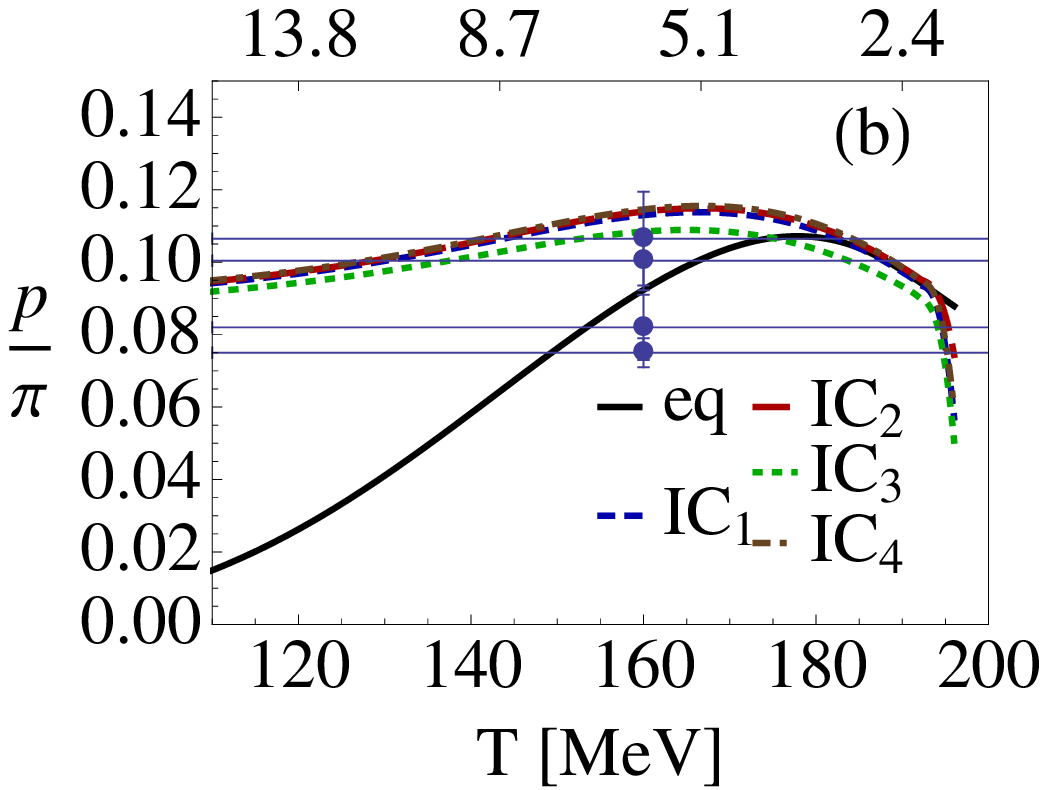,width=1\linewidth,clip=} 
\end{minipage}
\caption{ Results for the ratio of $p/\pi^-$ with various initial conditions for (a) $T_H=176$ MeV  or (b) $T_H=196$ MeV. Note that for STAR \cite{WhitePapers,STAR} $p/\pi^-0.11$ and  $\bar{p}/\pi^-=0.082$ and for PHENIX \cite{PHENIX} $p/\pi^-0.10$ ($p/\pi^+$ is actually measured but we convert it to $p/\pi^-$ to match STAR) and  $\bar{p}/\pi^-=0.047$ . } \label{fig:pppi}
\end{figure}

In our expanding fireball model we can calculate the $p/\pi$ ratio where the black solid line in each graph is the chemical equilibrium abundances, the colored lines are the dynamical calculations for various expansions that follow the calculated $T(t)$ relationship, and the error bars are the experimental data points. 
The pions, Hagedorn states, and $X\bar{X}$ all are allowed to chemical equilibrate, while we then vary the initial conditions and observe their effects.   Note that in all the following figures the effective numbers are shown so that the contribution of the Hagedorn states is included.  In Fig.\ \ref{fig:pppi} the ratio of $p/\pi^+$'s is shown.   We see that for $T_H=176$ MeV that our results enter the band of experimental data before $T=170$ MeV and remains there throughout the entire expansion regardless of the initial conditions. However, for  $T_H=196$ MeV the ratios already match the experimental data early on at $T\approx190$
 MeV.  They are briefly overpopulated around $T=160-170$ MeV but quickly return to the experimental values, except when the pions are overpopulated (this could imply that there are  too many Hagedorn states and a lower degeneracy of the Hagedorn states may produce better results).  
\begin{figure}
\begin{minipage}{0.5\linewidth}
\vspace{0pt}
\centering
\epsfig{file=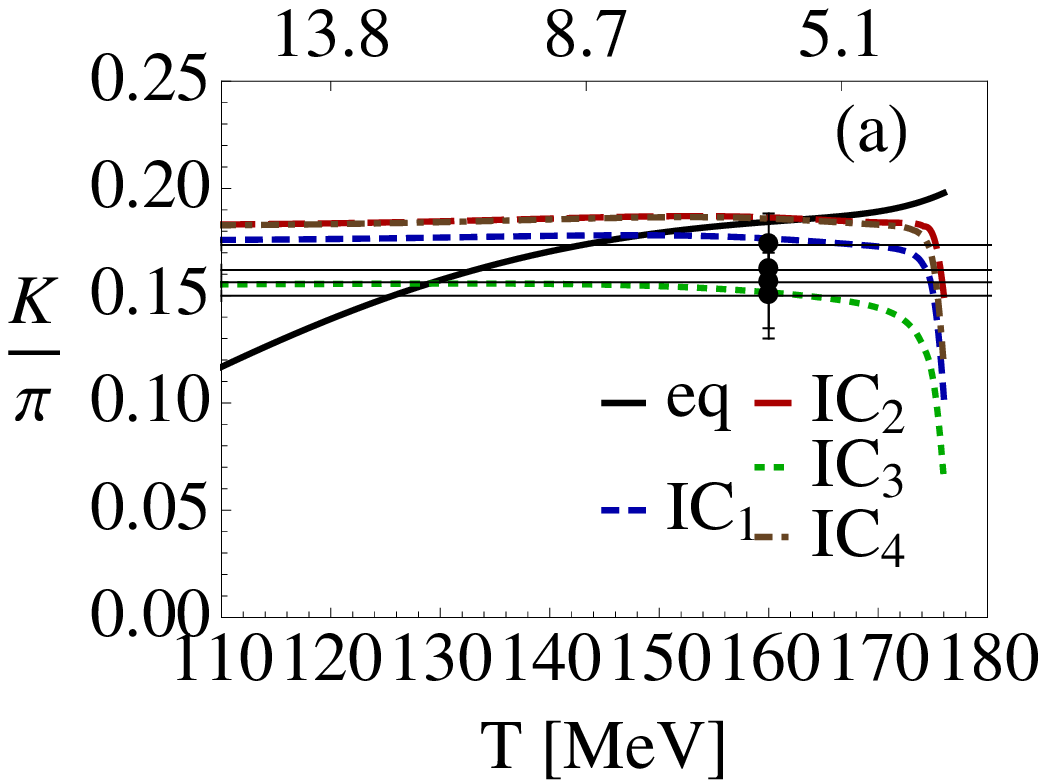,width=1\linewidth,clip=}
\end{minipage}
\hspace{0.5cm}
\begin{minipage}{0.5\linewidth}
\vspace{0pt}
\centering
\epsfig{file=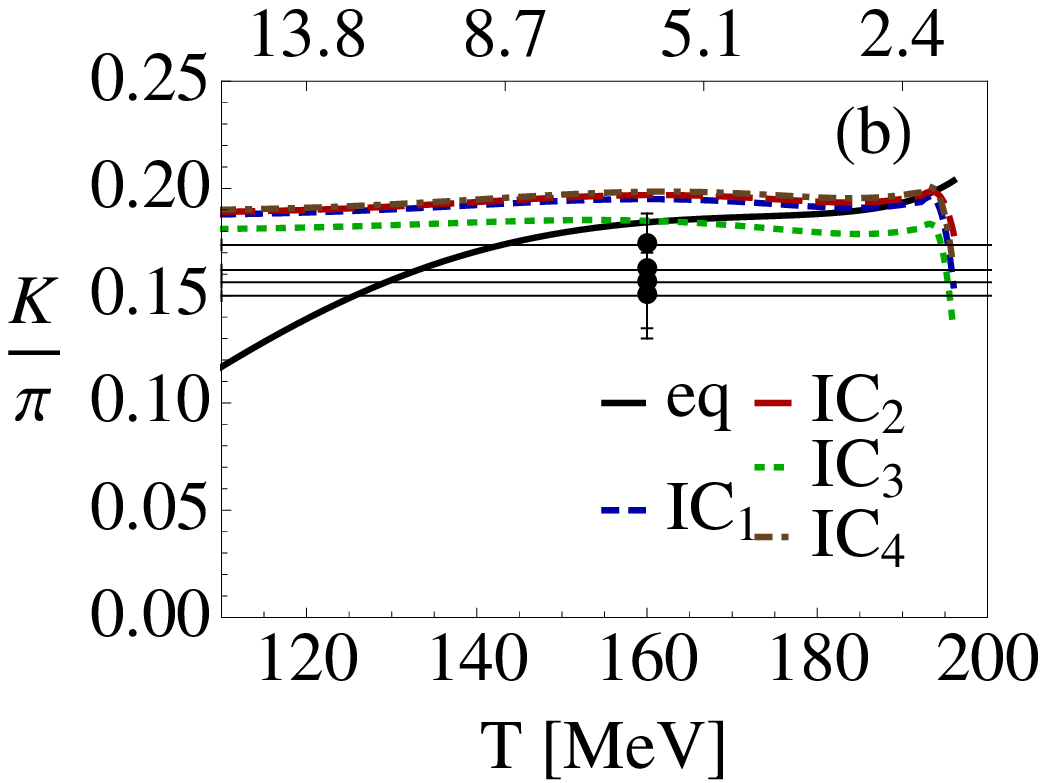,width=1\linewidth,clip=} 
\end{minipage}
\caption{ Results for the ratio of $K^+/\pi^-$ with various initial conditions for (a) $T_H=176$ MeV  or (b) $T_H=196$ MeV. Note that for STAR \cite{WhitePapers,STAR} $K^+/\pi^-=0.16$ and  $K^-/\pi^-=0.15$ and for PHENIX \cite{PHENIX} $K/\pi^-0.174$ ($K/\pi^+$ is actually measured but we convert it to $K/\pi^-$ to match STAR) and  $\bar{K}/\pi^-=0.162$ .  } \label{fig:KKpi}
\end{figure}

In Fig.\ \ref{fig:KKpi} the ratio of kaons to pions is shown for $T_H=176$ MeV and for $T_H=196$ MeV.   For $T_H=176$ MeV our results are roughly at the upper edge of the experimental values, while for $T_H=196$ MeV our results are slightly higher than the experimental values.  Regardless of initial conditions, the results at $T=110$ MeV are almost exactly those of the uppermost experimental data point.  
\begin{figure}
\begin{minipage}{0.5\linewidth}
\vspace{0pt}
\centering
\epsfig{file=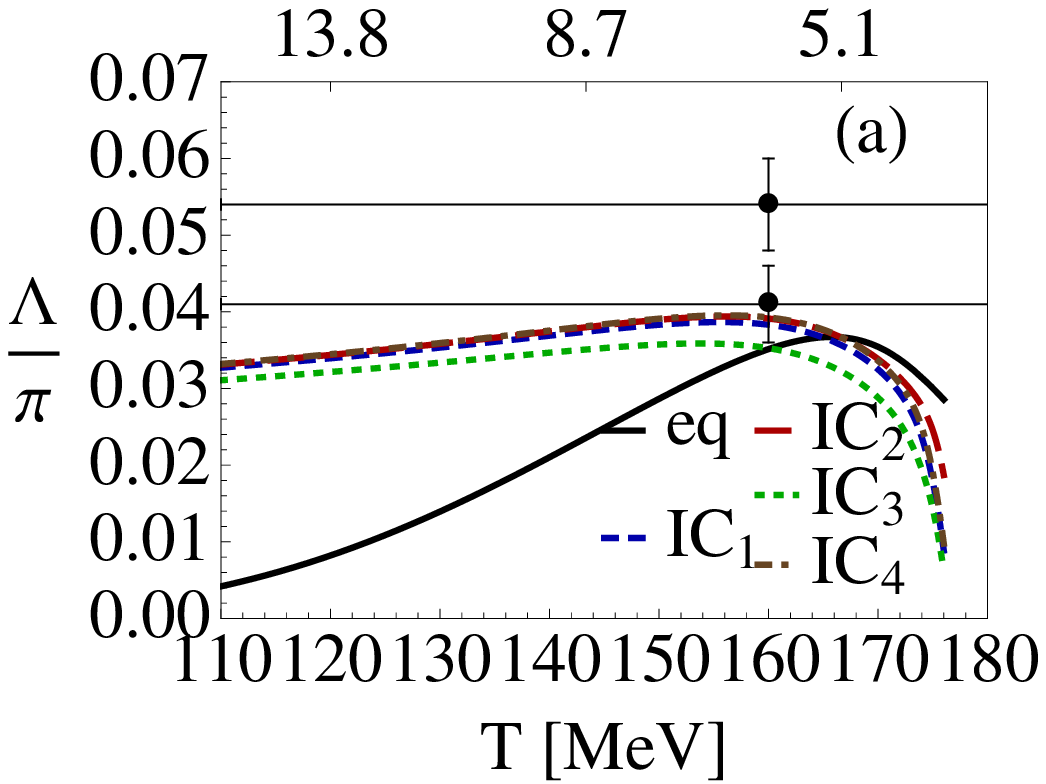,width=1\linewidth,clip=}
\end{minipage}
\hspace{0.5cm}
\begin{minipage}{0.5\linewidth}
\vspace{0pt}
\centering
\epsfig{file=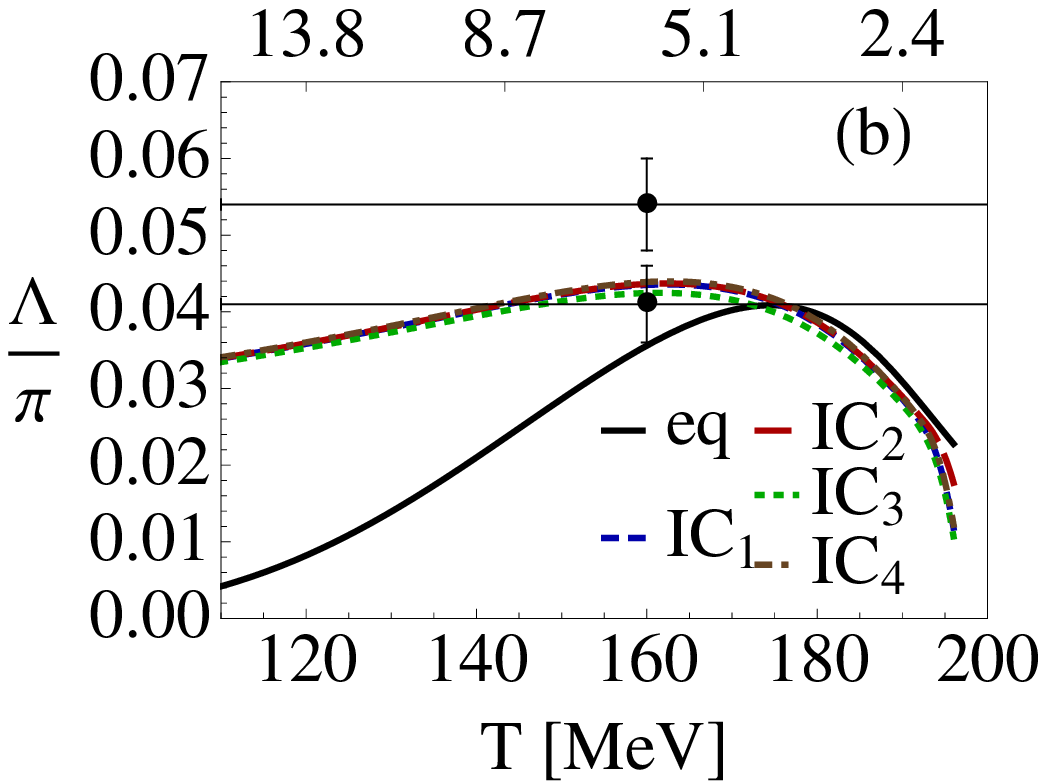,width=1\linewidth,clip=} 
\end{minipage}
\caption{ Results for the ratio of $\Lambda/\pi^-$'s with various initial conditions for (a) $T_H=176$ MeV  or (b) $T_H=196$ MeV. Note that for STAR \cite{WhitePapers,STAR} $\Lambda/\pi^-=0.54$ and  $\bar{\Lambda}/\pi^-=0.41$ and $\Lambda/\pi$ is not measured for PHENIX.} \label{fig:LLpi}
\end{figure}
The ratio of $\Lambda/\pi$'s is shown in Fig.\ \ref{fig:LLpi}. In both cases the $\Lambda/\pi$'s match the experimental values extremely well and the experimental values are reached already by $T\approx 170$ MeV.
\begin{figure}
\begin{minipage}{0.5\linewidth}
\vspace{0pt}
\centering
\epsfig{file=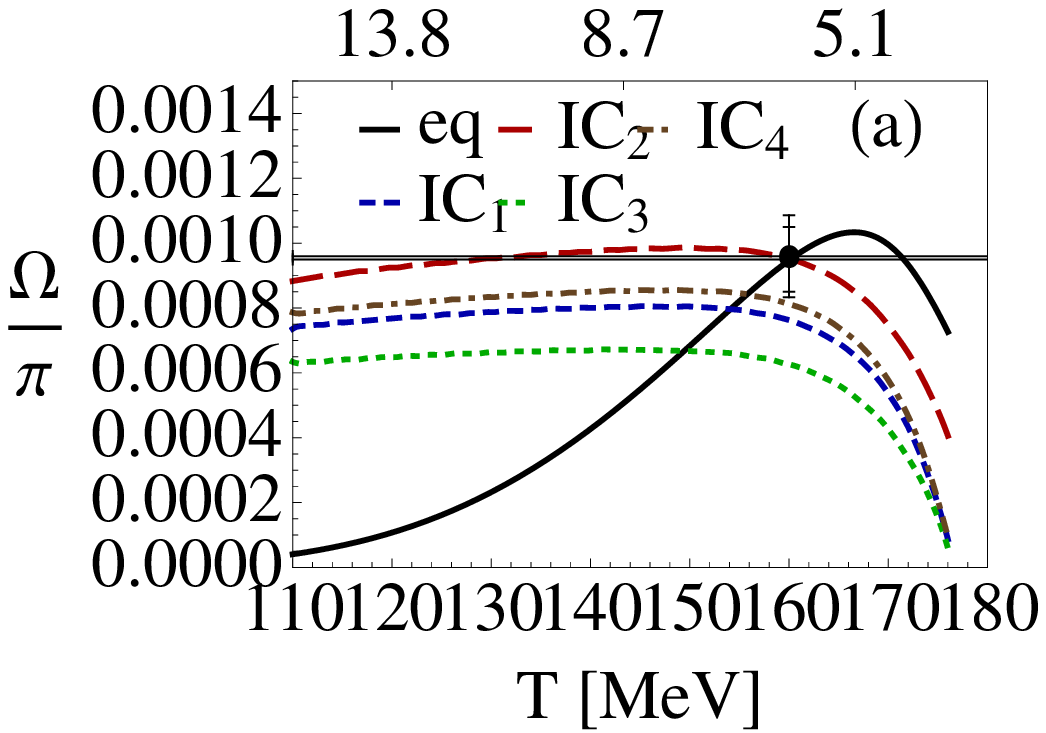,width=1\linewidth,clip=}
\end{minipage}
\hspace{0.5cm}
\begin{minipage}{0.5\linewidth}
\vspace{0pt}
\centering
\epsfig{file=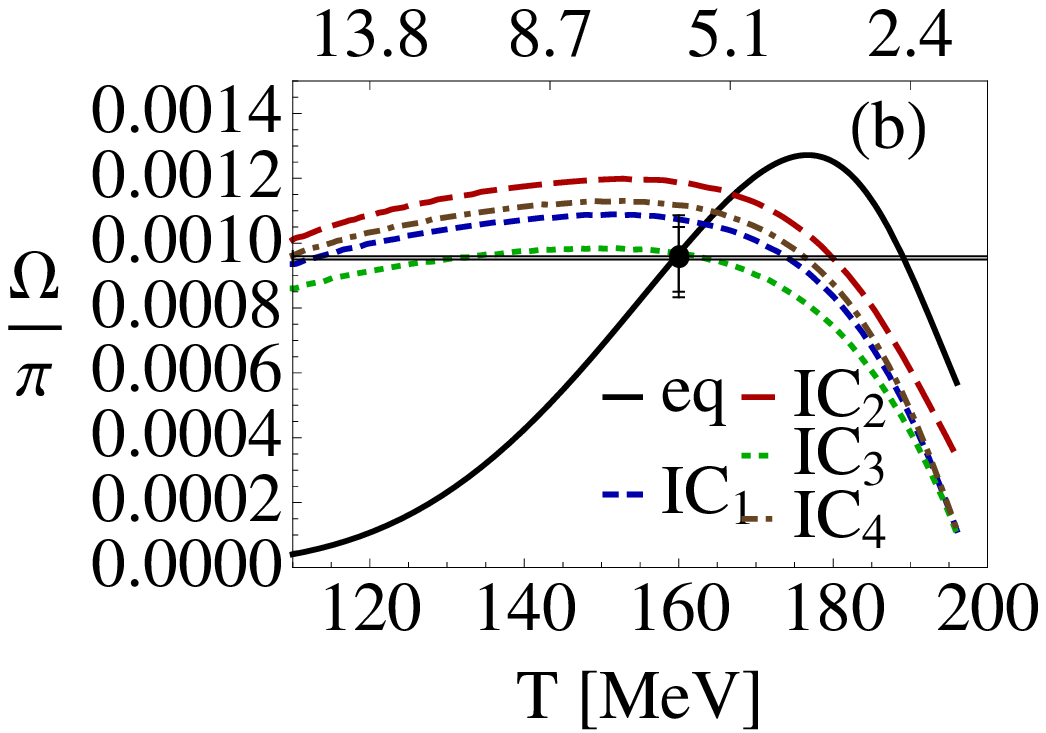,width=1\linewidth,clip=} 
\end{minipage}
\caption{ Results for the ratio of $\Omega/\pi^-$'s with various initial conditions for (a) $T_H=176$ MeV  or (b) $T_H=196$ MeV. Note that for STAR $\Omega/\pi^-=9.5 \times 10^{-4}$ and  $\bar{\Omega}/\pi^-=9.6 \times 10^{-4}$ STAR \cite{WhitePapers,STAR} and $\Omega/\pi$ is not measured for PHENIX. } \label{fig:OOpi}
\end{figure}
We can also use our model to investigate the possibility of $\Omega$'s.  In \cite{Greiner:2004vm}, they discussed the possibility of $\Omega$'s being produced using Hagedorn states. We are able to adequately populate the $\Omega\bar{\Omega}$ pairs so that they roughly match the experimental data.  
On the other hand, for the $\Omega $
particle the equilibration time is short only very close to $T_c$ (see \cite{NoronhaHostler:2009cf}).
The scenario is thus more delicate. If one would take, for example, one half the decay width of that of Eq. \ref{HSdecaywidth}  or one fourth of the decay width  the total production
of $\Omega $ is not sufficient up to 25 \%, or up to 50\%, respectively,  to meet the 
experimental yield (the other ratios are not significantly affected by such a change of
the decay width).  In a future work, it would be interesting to observe the other decay channels
that include exotic states, which also occur in the spirit
of Hagedorn states. In order to observe these decay channels 
a method, e.g. a microscopic quark model,  must be found to find the appropriate Hagedorn spectrum for strange mesonic/baryonic Hagedorn states. Branching ratios could be calculted using \cite{Max}.

\begin{figure}
\centering
\epsfig{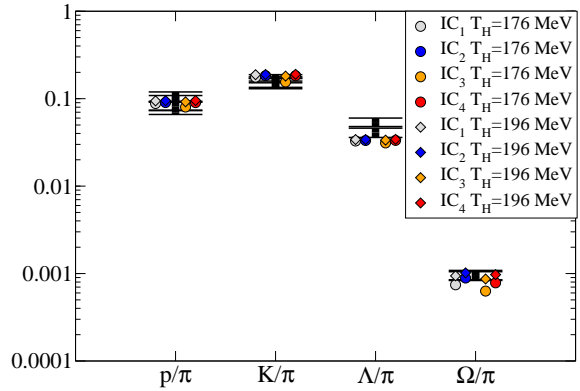} 
\caption{Plot of the various ratios including all initial conditions defined in Tab.\ \protect\ref{tab:IC}.  The points show the ratios at $T=110$ MeV for the various initial conditions (circles are for $T_H=176$ MeV and diamonds are for $T_H=196$ MeV).   The experimental results for STAR \protect\cite{WhitePapers,STAR} and PHENIX \protect\cite{PHENIX} are shown by the gray error bars.} \label{fig:summary}
\end{figure}

A summary graph of all the results of this section is shown in Fig.\ \ref{fig:summary}.  We see in our graph that our freezeout results match the experimental data well.
Thus, a dynamical scenario is able to explain chemical equilibration values that appear in thermal fits by $T=160$ MeV. In general, $T_H=176$ MeV and $T_H=196$ give chemical freeze-out values in the range between $T=160-170$ MeV.  Moreover, the initial conditions have little effect on the ratios and give a range in the chemical equilibrium temperature of about $\sim5$ MeV, which implies that information from the QGP regarding multiplicities is washed out due to the rapid dynamics of Hagedorn states.  Lower $\beta_i$ does slow the chemical equilibrium time slightly.  However, as seen in Fig.\ \ref{fig:summary} they still fit well within the experimental values.

\section{Transport Coefficients}

The large azimuthal asymmetry of low-$p_T$ particles and the strong quenching of high-$p_T$ probes measured at RHIC \cite{WhitePapers} indicate that the new state of matter produced in heavy ion collisions is a strongly interacting quark-gluon plasma \cite{Gyulassy:2004zy}. The matter formed in these collisions behaves almost as a perfect liquid \cite{etas} characterized by a very small value for its shear viscosity to entropy density ratio, which is in the ballpark of the lower bound $\eta/s \geq 1/(4\pi)$ \cite{Kovtun:2004de} derived within the anti-de Sitter/conformal field theory (AdS/CFT) correspondence \cite{Maldacena:1997re}. It was conjectured by Kovtun, Son, and Starinets (KSS) \cite{Kovtun:2004de} that this bound holds for all substances in nature (see, however, Refs.\ \cite{Cohen:2007qr,Dobado:2007tm} for possible counterexamples involving nonrelativistic systems).

Recent lattice calculations \cite{Meyer:2007ic} in pure glue $SU(3)$ gauge theory have shown that $\eta/s$ remains close to the KSS bound at temperatures not much larger than $T_c$. Additionally, calculations within the BAMPS parton cascade \cite{xucarsten}, which includes inelastic gluonic $gg\leftrightarrow ggg$ reactions, indicate that $\eta/s \sim 0.13$ in a purely partonic gluon gas \cite{Xu:2007ns}. Moreover, it was argued in \cite{Csernai:2006zz} that this ratio should have a minimum at (or near) the phase transition in quantum chromodynamics (QCD). This is expected because $\eta/s$ increases with decreasing $T$ in the hadronic phase \cite{prakash} (because the relevant hadronic cross section decreases with $T$) while asymptotic freedom dictates that $\eta/s$ increases with $T$ in the deconfined phase since in this case the coupling between the quarks and the gluons (and the transport cross section) descreases logarithmically \cite{amy}. Note, however, that in general perturbative calculations are not reliable close to $T_c$ (see, however, Ref.\ \cite{Hidaka:2008dr}).

Thus far, there have been several attempts to compute $\eta/s$ in the hadronic phase using hadrons and resonances \cite{Itakura:2007mx,Gorenstein:2007mw,otheretasHG}. However, these studies 
have not explicitly considered that the hadronic density of states in QCD is expected to be $\sim \exp(m/T_H)$ for sufficiently large $m$ \cite{Hagedorn:1968jf,exponentialspectrum}, where 
$T_H \sim 150-200$ MeV is the Hagedorn temperature.


Here a hadron resonance gas model which includes all the known particles and resonances with masses $m<2$ GeV \cite{Eidelman:2004wy} and 
also an exponentially increasing number of Hagedorn states (HS) \cite{NoronhaHostler:2007jf,NoronhaHostler:2009cf,thesis} is used to provide an 
upper limit on $\eta/s$ for hadronic matter close to the critical temperature that is comparable to $1/4\pi$. Additionally, we show that 
our model provides a good description of the recent lattice results \cite{Cheng:2007jq} for the speed of sound, $c_s$, close to $T_c=196$ MeV. 

\subsection{Results}

\begin{figure}[t]
\centering
\epsfig{file=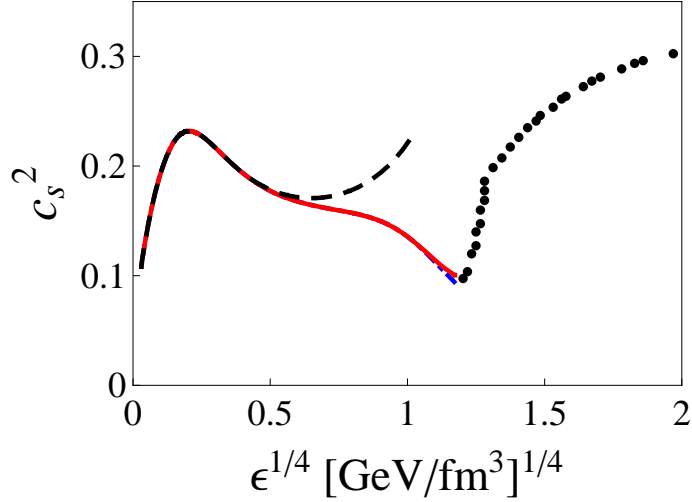,width=0.6\linewidth,clip=}
\caption{$c_s^2$ for the model including HS with $2<m<20$ GeV (solid red line) and for our hadron gas model that does not include HS (dashed black curve). The lattice results for the $p4$ action with $N_{\tau}=6$ \cite{Cheng:2007jq} are depicted in the dotted curve. } \label{fig:c_s}
\end{figure}

Using the thermodynamic model described in Sec. \ref{model}, we are able to study the effects of Hagedorn states on the transport coefficients of a hadron gas at high temperatures.  
In general, a very rapid increase in the number of particle species (specifically heavier species) around $T_c$ is expected to strongly reduce the speed of sound $c_s^2=dP/d\epsilon$ at the phase transition. While $c_s^2 \to 0$ at the transition would certainly lead to very interesting consequences for the evolution of the RHIC plasma \cite{Rischke:1996em}, recent lattice simulations have found that $c_s^2\simeq 0.09$ near $T_c$ \cite{Cheng:2007jq}. It is shown in Fig.\ \ref{fig:c_s} that $c_s^2(T\sim T_c)\sim 0.09$ in the model with HS while for the model without them $c_s^2 \sim 0.25$ near $T_c$. 


The total shear viscosity of our multi-component system computed within kinetic theory \cite{Reif} is $\eta_{tot} \sim \alpha \sum_i n_i \langle p_i \rangle \lambda_i$, where $n_{i}$ is the number density, $\langle p_{i} \rangle$ is the average momentum, and $\lambda_{i}$ is the mean free path for discrete states and HS ($\alpha\sim \mathcal{O}(1)$). Moreover, $\lambda_{i}=\left(\sum_{j} n_j\, \sigma_{ij}\right)^{-1}$, with $\sigma_{ij}$ being the scattering cross section. Due to their very large mass, the particle density of a HS is much smaller than that of discrete states. Thus, one can neglect the small contribution to the mean free path from terms involving the interaction between the standard hadrons and the HS. In this case, $\eta_{tot}=\eta_{HG}+\eta_{HS}$ where $\eta_{HG}$ is the shear viscosity computed using only the interactions between the standard hadrons while $\eta_{HS}=\frac{1}{3}\sum_{i}n_i \langle p\rangle_i \,\lambda_{i}$ includes only the contribution from HS, which move non-relativistically since $m_{HS}/T\gg 1$. Note that the approximation for $\eta_{tot}$ used here provides an upper bound for this quantity since the inclusion of the interactions between HS and the standard hadrons would only decrease $\eta_{tot}$. Using the results above, one sees that
\begin{eqnarray}
\label{etas1}
\left(\frac{\eta}{s}\right)_{tot}&\leq&\frac{s_{HG}}{s_{HG}+s_{HS}}\left[\left(\frac{\eta}{s}\right)_{HG}+ \frac{\eta_{HS}}{s_{HG}}\right].
\end{eqnarray}
While the entropy dependent prefactor in Eq.\ (\ref{etas1}) can be easily determined using our model, the detailed calculation of $\eta_{HG}$ and $\eta_{HS}$ requires the knowledge about the mean free paths of the different particles and resonances in the thermal medium. In the non-relativistic approximation, we can set $\langle p_i \rangle = m_i \langle v_i \rangle = \sqrt{8m_i\,T/\pi}$ in Eq.\ (\ref{etas1}). Note that HS with very large $m_i$'s are more likely to quickly decay. We assume that $\lambda_i=\tau_i \,\langle v_i\rangle$ where $\tau_i\equiv 1/\Gamma_i=1/(0.151\, m_i-0.0583)$ GeV$^{-1}$ is the inverse of the decay width of the $i^{th}$ HS obtained from a linear fit to the decay widths of the known resonances in the particle data book \cite{NoronhaHostler:2007jf,NoronhaHostler:2009cf,Senda,thesis}. Our choice for $\lambda_i$ gives the largest mean free path associated with a given state because it neglects any possible collisions that could occur before it decays on its own. Note, however, that the decay cross section is in general different than the relevant collision cross section for momentum transport that contributes to $\eta$ according to kinetic theory. Thus, it is not guaranteed a priori that these decay processes contribute to $\eta$ in the usual way.

\begin{figure}[t]
\centering
\epsfig{file=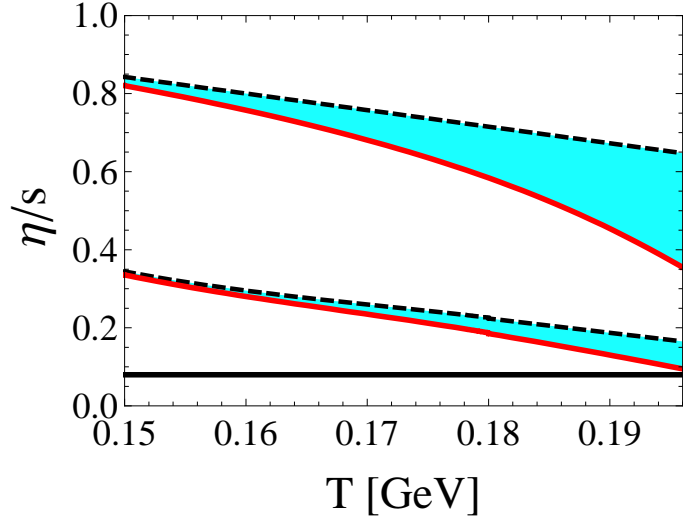,width=0.6\linewidth,clip=}
\caption{$\eta/s$ is shown for a gas of pions and nucleons \cite{Itakura:2007mx} (upper dashed black line) and for a hadron resonance gas with (constant) excluded volume corrections \cite{Gorenstein:2007mw} (lower dashed black line). An upper bound on the effects of HS on $\eta/s$ is shown in solid red lines. The blue band between the curves is used to emphasize the effects of HS. The solid black line at the bottom is the AdS/CFT lower bound $\eta/s=1/4\pi$ \cite{Kovtun:2004de}.} \label{fig:eta_s}
\end{figure}

We find that $\eta_{HS} =8T\sum_{i}\,n_{i}\tau_i/3\pi$. The remaining ratio $\left(\eta/s\right)_{HG}$ has been computed in \cite{Itakura:2007mx,Gorenstein:2007mw,otheretasHG} using different models and approximations. Since our main goal is to understand the effects of HS on  $(\eta/s)_{tot}$, here we will simply use the values for $\left(\eta/s\right)_{HG}$ obtained in some of these calculations to illustrate the importance of HS. We chose to obtain $(\eta/s)_{HG}$ for a gas of pions and nucleons from \cite{Itakura:2007mx} and for a hadron resonance gas from \cite{Gorenstein:2007mw}. Note that the results for $\eta/s$ obtained from the calculation that included many particles and resonances \cite{Gorenstein:2007mw} are already much smaller than those found in \cite{Itakura:2007mx}. A linear extrapolation of the results in \cite{Itakura:2007mx,Gorenstein:2007mw} was used to obtain their $\eta/s$ values at high temperatures. One can see in Fig.\ \ref{fig:eta_s} that $\left(\eta/s\right)_{tot}$ drops significantly around $T_{c}$ because of HS.  This result is especially interesting because $\eta/s$ in the hadronic phase is generally thought to be a few times larger than the string theory bound. One can see that the contributions from HS should lower $\eta/s$ to near the KSS bound close to $T_c$. Thus, the drop in $\eta/s$ due to HS could explain the low shear viscosity near $T_c$ already in the hadronic phase. 

\section{Thermal Fits}

Thermal fits computed within grand canonical statistical models are normally used to reproduce hadron yield ratios in heavy ion collisions \cite{StatModel,Schenke:2003mj,Andronic:2005yp,Manninen:2008mg,thermalmodels,RHIC}.  Thermal models computed at AGS, SIS, SPS, and RHIC energies can be used to construct a chemical freeze-out line in the QCD phase diagram \cite{Cleymans:2007kj,freezeoutline}.  For Au+Au collisions at RHIC$\;$ at $\sqrt{s_{NN}}=200$ GeV, specifically, estimates for the chemical freeze-out temperature and baryon chemical potential range from $T_{ch}=155-169$ MeV and $\mu_b=20-30$ MeV \cite{Andronic:2005yp,Manninen:2008mg,RHIC}. These thermal models give reasonable fits to the experimental data, which leads to the conclusion that chemical freeze-out may be reached in these experiments. Since Hagedorn states have been shown to affect the chemical equilibration times, thermodynamic properties, and transport coefficients of hadron resonance gases close to $T_c$ it is natural to expect that they may also be relevant in the thermal description of particle ratios.

\section{Model}

In order to calculate the baryonic chemical potential $\mu_b$ and the strange chemical potential $\mu_s$ we use the following conservation relation
\begin{eqnarray}\label{eqn:cons}
0&=&\frac{\sum_i n_i S_i}{\sum_i n_i B_i},
\end{eqnarray}
so the total strangeness per baryon number is held at zero. There $n_i$ is the density of the $i^{th}$ particle that has a corresponding baryon number $B_i$ and strangeness $S_i$. The Hagedorn states are implemented in our model as previously discussed through the effective numbers.   It is important to note here that because the Hagedorn states always produce pairs of $X\bar{X}$'s that the entire contribution  to ratios like $K^+/K^-$ must come from the known particles.  Therefore, in our calculations the baryonic chemical potential, which is directly related to the strange chemical potential, is somewhat inflated.  If we were to include baryonic and/or strange Hagedorn states then $\mu_b$ would be lower. 
In order to get an idea of the quality of the thermal fits, we define $\chi^2$ as
\begin{equation}
\chi^2=\sum_i \frac{\left(R_i^{exp}-R_i^{therm}\right)^2}{\sigma_i^2}
\end{equation}
where $R_i^{therm}$ is our ratio of hadron yields calculated within our thermal model whereas $R_i^{exp}$ is the experimentally measured value of the hadron yield with its corresponding error $\sigma_i^2$.  

In this work we look at only the experimental values at mid-rapidity and we used only the systematic error given by each respective experiment. We vary the temperature and $\mu_b$ according to the conservation laws in Eq.\ (\ref{eqn:cons}) to minimize $\chi^2$.  We use data from STAR \cite{STAR} and PHENIX \cite{PHENIX} for Au+Au collisions at RHIC at $\sqrt{s_{NN}}$=200 GeV and observe: $\pi^{-}/\pi^{+}$, $\bar{p}/p$, $K^-/K^+$, $K^+/\pi^+$, $p/\pi^+$, and $(\Lambda+\bar{\Lambda})/\pi^+$.  All are calculated by STAR \cite{STAR}, however, only $\pi^{-}/\pi^{+}$, $\bar{p}/p$, $K^-/K^+$, $K^+/\pi^+$, $p/\pi^+$ are given by PHENIX.  Because there is such a difference between $p/\pi^+$ from PHENIX and STAR we choose only the value from STAR so that we can compare are results to \cite{Andronic:2005yp} where they also exclude $p/\pi^+$ from PHENIX.  It should be noted that \cite{Andronic:2005yp} includes more ratios than we do such as multi-strange particles and resonances.  The purpose of the present study is not to confirm their results but to compare thermal fits that with and without Hagedorn states.

\section{Results}

\begin{figure}
\begin{minipage}{0.5\linewidth}
\vspace{0pt}
\centering
\epsfig{file=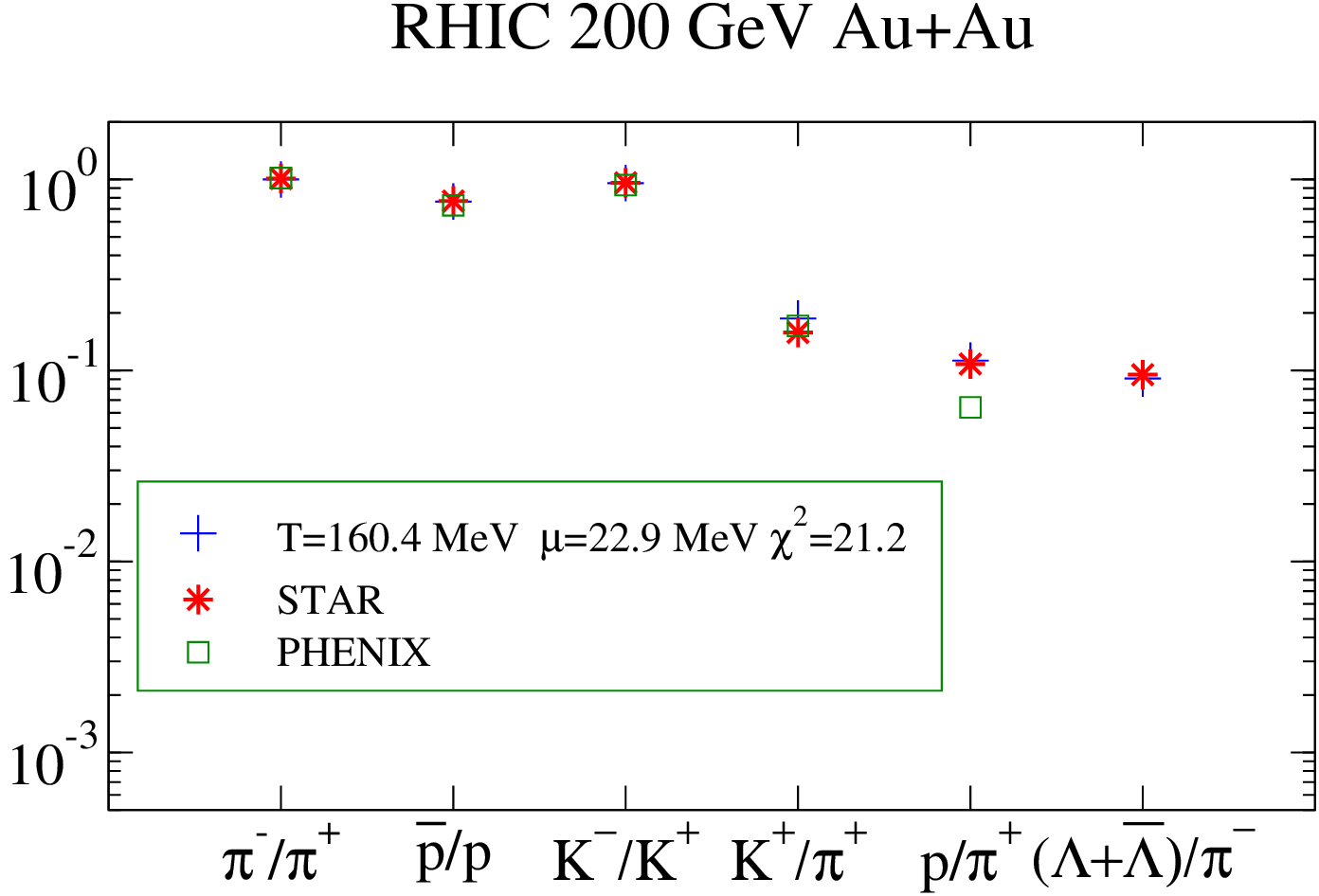,width=0.5\linewidth,clip=}
\caption{Thermal fits for Au+Au collisions at RHIC at $\sqrt{s_{NN}}=200$ GeV when no Hagedorn states are present.  }
\label{fig:noHS}
\end{minipage}
\hspace{0.5cm}
\begin{minipage}{0.5\linewidth}
\vspace{0pt}
\centering
\epsfig{file=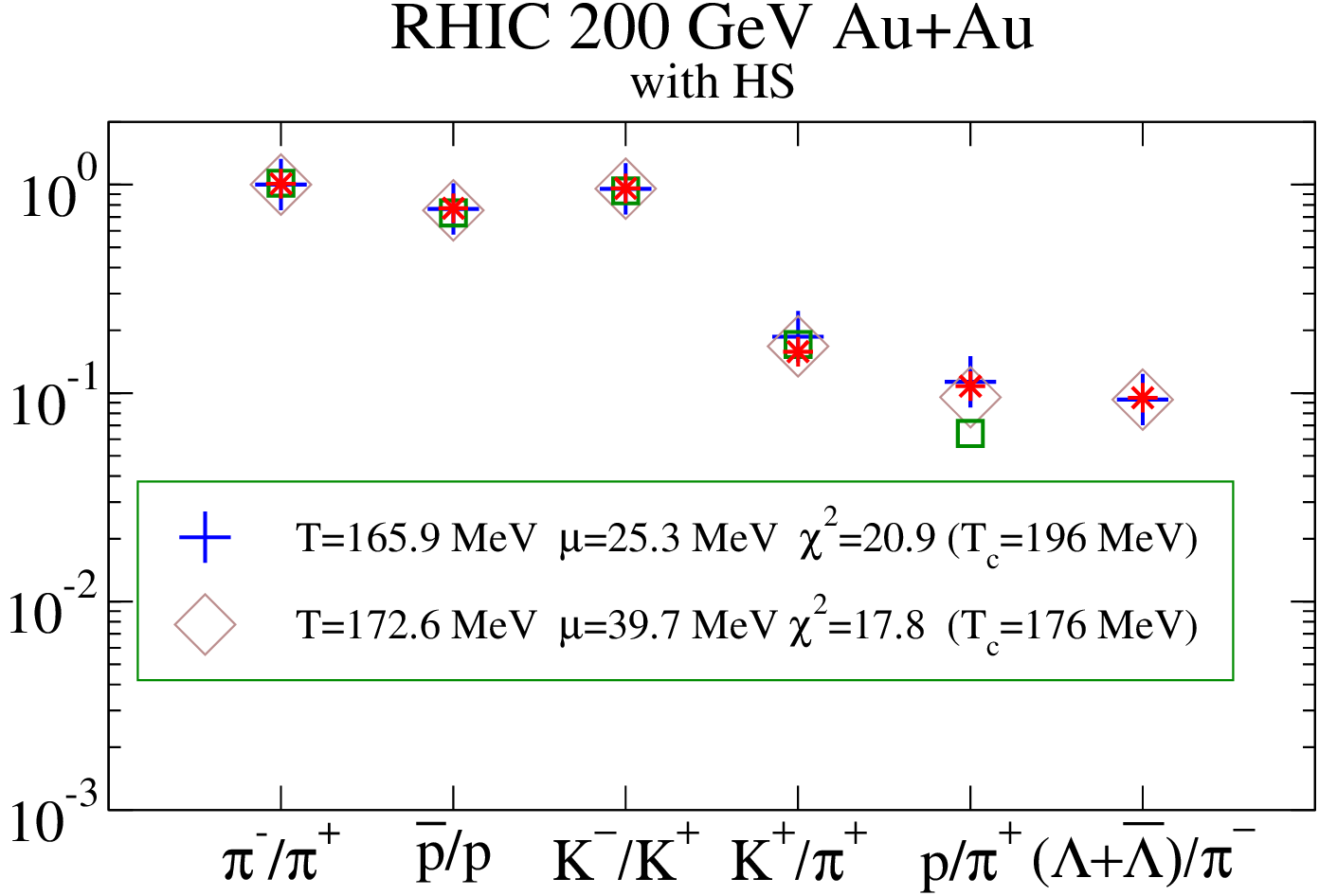,width=0.5\linewidth,clip=}
\caption{Thermal fits including Hagedorn states for Au+Au collisions at RHIC at $\sqrt{s_{NN}}=200$ GeV.  }
\label{fig:HS}
\end{minipage}
\end{figure}

For a hadron gas excluding Hagedorn states (see Fig.\ \ref{fig:noHS}),  $T_{ch}=160.4$ MeV and $\mu_b=22.9$ MeV, which gave $\chi^2=21.2$.  Our resulting temperature and baryonic chemical potential are almost identical to that in \cite{Andronic:2005yp}. 
The inclusion of Hagedorn states is our primary interest.  Starting with the fit for the RBC-Bielefeld collaboration, we obtain $T_{ch}=165.9$ MeV, $\mu_b=25.3$ MeV, and $\chi^2=20.9$, which is shown in Fig.\ \ref{fig:HS}.  The $\chi^2$ is slightly smaller than in Fig.\ \ref{fig:noHS}.
When we consider the lattice results from BMW where $T_c=176$ MeV, we find $T_{ch}=172.6$ MeV, $\mu_b=39.7$ MeV, and $\chi^2=17.8$. The lower critical temperature seems to have an impact on the thermal fit. The lower $\chi^2$ is due to the larger contribution of Hagedorn states at at $T_{ch}=172.6$ MeV, which is much closer to $T_c$.  The contributions of the Hagedorn states to the total number of the various species at this temperature and chemical potential are about $30-50\%$.
The difference in the $\chi^2$'s for BMW and RBC-Bielefeld collaboration is directly related to the contribution of Hagedorn states in the model.  Because the RBC-Bielefeld critical temperature region is significantly higher than its corresponding chemical freeze-out temperature the contribution of the Hagedorn states is minimal at only 4-11$\%$.  
We find that the inclusion of Hagedorn states should not only provide a better fit but they also affect the chemical freeze-out temperature and the baryonic chemical potential. The more mesonic Hagedorn states are present the larger $\mu_b$ becomes.  Furthermore, our fits also have higher $T_{ch}$'s than seen in the fit without the effects of Hagedorn states.

\section{Conclusions}

Dynamical reactions with the known hadronic particles cannot account for the particle abundances seen at RHIC. The chemical equilibration times are too long and do not fit within the calculated time scale of the hadronic fireball.  This has led to the assumption that the chemical freeze-out temperature and the critical temperature coincide.  However, assuming that these heavy, quickly decaying Hagedorn states exist, chemical equilibrium can be achieved on short enough time scales that fit within a hadronic, cooling fireball i.e. on the order of $\approx 1-2\;fm/c$. Moreover, Hagedorn states states provide a very efficient way for incorporating multi-hadronic
interactions (with parton rearrangements). This work indicates that the population and repopulation of potential
Hagedorn states close to phase boundary can be the key source for a dynamical understanding of generating and
chemically equilibrating the standard and measured hadrons.  

Because of the success of decays from Hagedorn states in reproducing experimental particle ratios, it was only logical to extend their use to other areas.  Hadronic models with the known particles are not able to reproduce the low shear viscosity to entropy density ratio that seems to be required to explain the large elliptic flow observed at RHIC. One might expect that by increasing the number of particles in a gas the mean free path would subsequently decrease, which would in effect decrease the total $\eta/s$. In fact, including the ``missing" Hagedorn states also decreases $\eta/s$ for hadronic matter near $T_c$ near to the string theory value $1/(4\pi)$. Moreover, according to the general argument that small $\eta/s$ implies strong jet quenching \cite{Majumder:2007zh}, the significant reduction of $\eta/s$ indicates that hadronic matter near the phase transition is more opaque to jets than previously thought. Since the system should spend most of its time near $T_c$ (because of the minimum in the speed of sound), the fact that $\eta/s$ can be very small in that region in the hadronic phase may imply that the key observables for the QGP, i.e., the strong quenching of jets and the large elliptic flow, can receive significant contributions from the hot Hagedorn resonance gas. 

We assumed that the particle ratios measured in Au+Au collisions at RHIC at $\sqrt{s_{NN}}=200$ GeV admit a purely statistical description at chemical freeze-out. Our results for thermal fits without Hagedorn states concur well with other thermal fit models \cite{Andronic:2005yp} where the chemical freeze-out temperature ($T_{ch}=160.4$ MeV)  is almost identical and the baryonic chemical potential ($\mu_b=22.9$ MeV)  is only slightly larger. The thermal fit with the known particles in the particle data group provides a decent fit with $\chi^2=21.2$.  However, the inclusion of Hagedorn states provides an even better fit to the experimental data where $\chi^2=17.8$ for the BMW collaboration and $\chi^2=20.9$ for the RBC-Bielefeld collaboration.  This provides further evidence \cite{NoronhaHostler:2007jf,NoronhaHostler:2009cf,NoronhaHostler:2007fg,NoronhaHostler:2008ju} that Hagedorn states should be included in a description of hadronic matter near $T_c$.

\section*{Acknowledgements}
This work has been supported by the Helmholtz International Center for FAIR within the framework of the
LOEWE program launched by the State of Hesse.

\end{document}